\documentclass[11pt]{article}
\pdfoutput=1
\usepackage{soul}
\usepackage{jheppub}
\usepackage{amsfonts}
\usepackage{array}
\usepackage{amsmath}
\allowdisplaybreaks[4]         
\usepackage{amssymb}   
\usepackage{euscript}       
\usepackage{xcolor}          
\usepackage{tensor}     
\usepackage{caption}
\usepackage{subcaption}   
\usepackage{graphicx}
\usepackage[T1]{fontenc} % if needed
\usepackage{float}
\newcommand{\req}[1]{(\ref{#1})} %{Eq.\thinspace(\ref{#1})}  

\newcommand{\bea}{\begin{eqnarray}}
	\newcommand{\eea}{\end{eqnarray}}
\newcommand{\ba}{\begin{eqnarray}}
	\usepackage{braket}
	\newcommand{\ea}{\end{eqnarray}}

\newcommand{\beq}{\begin{equation}}
	\newcommand{\eeq}{\end{equation} }
\newcommand{\beqa}{\begin{eqnarray}}
	
	\newcommand{\eeqa}{\end{eqnarray}}
\newcommand{\beqar}{\begin{eqnarray*}}
	\newcommand{\eeqar}{\end{eqnarray*}}

\newcommand{\be}{\begin{equation}}
	\newcommand{\ee}{\end{equation}}
\renewcommand{\req}[1]{(\ref{#1})}

\newcommand{\E}{\mathcal{E}}

\allowdisplaybreaks

\title{Eikonal quasinormal modes of highly-spinning black holes in higher-curvature gravity: a window into extremality}

\author[a,b]{Pablo A. Cano}
\author[c,d]{, Marina David}
\author[b]{and Guido van der Velde}

\affiliation[a]{Departamento de F\'isica, Universidad de Murcia, Campus de Espinardo, 30100 Murcia, Spain}

\affiliation[b]{Departament de F\'isica Qu\`antica i Astrof\'isica, Institut de Ci\`encies del Cosmos Universitat de Barcelona, Mart\'i i Franqu\`es 1, E-08028 Barcelona, Spain}

\affiliation[c]{Instituut voor Theoretische Fysica, KU Leuven.
	Celestijnenlaan 200D, B-3001 Leuven, Belgium \vspace{0.1cm}}
\affiliation[d]{Leuven Gravity Institute, KU Leuven, Celestijnenlaan 200D, B-3001 Leuven, Belgium}

\emailAdd{pablocano@um.es}
\emailAdd{marina.david@kuleuven.be}
\emailAdd{guidovandervelde@icc.ub.edu}
\date{\today}
\abstract{We carry out the first computation of gravitational quasinormal modes of black holes with arbitrary rotation in a theory with higher-derivative corrections. Our analysis focuses on a recently identified quartic-curvature theory that preserves the isospectrality of quasinormal modes in the eikonal limit and that is connected to string theory. We find a master equation that governs large-momentum gravitational perturbations in this theory. By solving this equation with WKB methods, we provide complete results for the corrections to the Kerr quasinormal mode frequencies for arbitrary spin and arbitrary $\mu=m/(\ell+1/2)$, where $\ell$ and $m$ are the harmonic numbers.  Our results show that the corrections become orders of magnitude larger when the spin is close to extremality, with the modes close to the critical value of $\mu$ that separates damped and zero-damped modes being particularly sensitive.  We also perform a geometric-optics analysis of gravitational-wave propagation around black holes and relate the equatorial ``graviton-sphere'' orbits to quasinormal mode frequencies with $\ell=m$. We find that the usual correspondence between the Lyapunov exponent of those orbits and the imaginary part of the frequency is modified.

}

\begin{document} 
	\maketitle
	\flushbottom
	%%%%%%%%%%%%%%%%%%%%%%%%%%%%%%%%%%%%%
	%\setcounter{tocdepth}{2}
	%the line above sets the depth of the table of contents. {2} means it will display section and subsections only.
	%{\small
		%\setlength\parskip{-0.5mm} 
		%\tableofcontents
		%}

	\newpage
	\allowdisplaybreaks

\section{Introduction}
The late-time response of a perturbed black hole is governed by a set of damped sinusoidal modes known as quasinormal modes (QNMs). According to General Relativity (GR), the QNM spectrum of a black hole is determined solely by its mass and spin\footnote{We consider vacuum GR so we do not discuss the case of electric charge.}, on account of the no-hair theorems. Consequently, the entire spectrum of a black hole can be predicted from the measurement of a single mode, and the detection of additional modes provides consistency checks. This simplicity makes black hole spectroscopy one of the most promising approaches to test GR with current and future gravitational-wave (GW) detectors \cite{Kokkotas:1999bd,Dreyer:2003bv,Berti:2009kk,Konoplya:2011qq,Berti:2018cxi,Berti:2018vdi,Barack:2018yly,Barausse:2020rsu,Berti:2025hly}. 

Black hole spectroscopy also gives us a unique opportunity to test modifications of GR. While theories beyond GR can still preserve the no-hair theorems and the uniqueness of the spectrum, they affect the gravitational dynamics and lead to modifications of the QNM frequencies. Therefore, by determining the QNMs in ringdown observations we can distinguish GR from its alternatives and set constraints on specific modifications of Einstein's theory \cite{Silva:2022srr,2023arXiv230501696F,Maselli:2023khq,Liu:2024atc,Maenaut:2024oci}. Motivated by this perspective, the analysis of black hole QNMs in modified gravity  has become a very active area of research in recent years --- see \cite{Cardoso:2009pk,Blazquez-Salcedo:2016enn,Cardoso:2018ptl,deRham:2020ejn,Konoplya:2020bxa, Cano:2020cao,Moura:2021nuh,Moura:2021eln,Pierini:2021jxd,Wagle:2021tam,Srivastava:2021imr,Bryant:2021xdh,Cano:2021myl,Pierini:2022eim,Cano:2023jbk,Miguel:2023rzp,Mukohyama:2023xyf,Silva:2024ffz,Konoplya:2024hfg,Chung:2024ira,Chung:2024vaf,Blazquez-Salcedo:2024oek,Cano:2024ezp,Khoo:2024agm,Antoniou:2024jku,Blazquez-Salcedo:2024dur,Li:2025fci,Lestingi:2025jyb,Boyce:2025fpr} for a necessarily incomplete list.  

In order to test particular theories via black hole spectroscopy, the theoretical modeling is indeed crucial: the spectrum of QNMs needs to be determined for each theory. Furthermore, this should be done for the astrophysically relevant case in which the black hole rotates.  The computation of QNMs of rotating black holes beyond GR is a remarkably challenging problem and only recently significant progress has been achieved. 
The introduction of modified Teukolsky equations \cite{Li:2022pcy,Hussain:2022ins,Cano:2023tmv,Cano:2024bhh}
, and independently, the development of spectral methods applied to the (modified) Einstein's equations \cite{Chung:2023zdq,Blazquez-Salcedo:2023hwg,Chung:2023wkd}, has led to the first results of QNMs of black holes with relatively large angular momentum in several theories beyond GR \cite{Cano:2023jbk,Chung:2024ira,Chung:2024vaf,Blazquez-Salcedo:2024oek,Cano:2024ezp,Khoo:2024agm,Blazquez-Salcedo:2024dur,Chung:2025gyg}. 

Despite this remarkable progress, we are still far from achieving a complete understanding of the QNM spectrum of rotating black holes in these theories.  Neither the highly damped modes, the eikonal regime or the case of highly-rotating (near-extremal) black holes\footnote{The case of near-extremal, non-rotating charged black holes has been recently addressed in \cite{Boyce:2025fpr}.} have been thoroughly analyzed. In this article, we tackle the two latter questions simultaneously.

Understanding the spectrum of highly rotating black holes is undoubtedly the most important question, as there are reasons to suspect that it can be particularly sensitive to the effects of new physics. Notably, the spectrum of near-extremal Kerr black holes is known to possess a rich behavior with infinitely many modes becoming long-lived  \cite{Hod:2008zz,Hod:2009td,Hod:2012bw,Yang:2012pj,Yang:2013uba}. Since this is related to some QNM frequencies having vanishing imaginary part, it is natural to expect that these could be particularly sensitive to  corrections. In a related context, it has been recently observed by \cite{Horowitz:2023xyl}  that higher-derivative corrections can lead to the formation of singularities on the horizon of extremal Kerr black holes --- see also \cite{Horowitz:2024dch,Cano:2024bhh,Chen:2024sgx}. This shows that non-trivial effects can appear when we consider the extremal limit in the presence of higher-derivative corrections. 

In order to make progress in this direction, here we focus on a particular higher-curvature extension of GR identified in \cite{Cano:2024wzo}. Ref.~\cite{Cano:2024wzo} considered a general effective-field-theory (EFT) extension of GR and showed that, to eight-derivative order, there is a unique higher-curvature term that leads to non-birefringent GW propagation on curved backgrounds in the large momentum limit.  It was proven that such theory preserves the isospectrality of QNMs in the eikonal limit in the case of static black holes, and it was conjectured that the same property would hold for rotating black holes --- here we confirm this statement. On a physical level, this theory can be motivated by the idea of preserving natural properties of GR --- in particular, non-birefringence can be understood as a symmetry between the two degrees of freedom of the graviton. In addition, it turns out that the theory coincides with the leading-order higher-curvature correction predicted by type II string theories \cite{Gross:1986iv}. 
On a practical level, the fact that the there is a single dispersion relation for both polarization modes of the graviton implies a simplification in the analysis of gravitational perturbations of large momentum. In this paper, we exploit this property in order to obtain the eikonal QNMs for black holes with arbitrary rotation. 

Eikonal QNMs are interesting on their own due to their geometric correspondence with photon-sphere geodesics around the black hole \cite{Ferrari:1984zz,Cardoso:2008bp, Hod:2009td, Dolan:2011fh,Hod:2012bw,Yang:2012he}, allowing one to obtain analytic expressions for the QNM frequencies and related quantities \cite{Konoplya:2023moy,Konoplya:2024lir,Chen:2024hum}. 
A generalization of this correspondence --- with the photon-sphere replaced by a ``graviton-sphere'' \cite{Papallo:2015rna} --- is expected to hold in modified theories of gravity \cite{Glampedakis:2019dqh,Silva:2019scu,Bryant:2021xdh,Miguel:2023rzp,Chowdhury:2024uzd}. Most of the current analyses in this direction have focused on the case of spherically symmetric black holes. As a notable exception, Ref.~\cite{Miguel:2023rzp} has obtained the eikonal QNMs of electromagnetic perturbations of Kerr black holes in a theory with non-minimal couplings between the Maxwell field and the curvature. To the best of our knowledge, an analogous computation in the case of gravitational perturbations in higher-derivative gravity has not appeared yet in the literature.

The paper is organized as follows
\begin{itemize}
\item In section~\ref{sec:theory} we review the isospectral EFT of \cite{Cano:2024wzo} and we study the linearized equations on a curved background. We show that gravitational perturbations of large momentum can be described by a master scalar equation containing higher-derivative corrections to the wave operator.  
\item In section~\ref{sec:geometricoptics} we consider the geometric optics limit of the master equation and analyze the circular orbits followed by GWs on the equatorial plane of a rotating black hole. We link the properties of these orbits to  the QNMs with $\ell=m$. We show that the orbital frequency is still associated with the real part of QNM frequencies, but the imaginary part is no longer proportional to the Lyapunov exponent associated to the instability timescale of individual orbits. We obtain instead the correct values of the imaginary part by analyzing the decay timescale of a bundle of equatorial GW orbits. 
%We link the frequency of these orbits to the real part of QNM frequencies and we also relate the damping time of a bundle of such orbits with the imaginary part, obtaining analytic expressions in both cases. We show that the damping factor is no longer proportional to the Lyapunov exponent associated to the instability timescale of individual orbits. 
\item In section~\ref{sec:effective}, we directly solve the master scalar equation by decomposing it into spheroidal harmonics and finding a modified Teukolsky radial equation. We find an exact expression for the modification of the Teukolsky potential for all values of the black hole rotation and of the ratio $\mu=m/(\ell+1/2)$, where $\ell$ and $m$ are harmonic numbers.  We then solve for the QNMs by employing the WKB method. 
\item In section~\ref{sec:QNMs}, we present detailed results for the corrections to the Kerr QNM frequencies as a function of the black hole rotation and of $\mu$. We check that, for moderate rotation, the exact QNMs obtained from the Teukolsky equation \cite{Cano:2023jbk,Cano:2024ezp} are well approximated by our eikonal results as we increase $\ell$.  We then analyze the behavior of the eikonal modes for high rotation, finding that the corrections are generically much larger near extremality. We uncover that modes around the critical value $\mu_{\rm cr}\approx 0.744$ --- that separates damped modes from zero-damped modes --- are particularly sensitive to the corrections, as their lifetimes could potentially be dramatically affected. We discuss the possibility of the corrections becoming of order one within the regime of validity of the EFT approach. 

\item We present our conclusions and discuss some future directions in section~\ref{sec:conclusions}.
\end{itemize}

 %%%%%%%%%%%%%%%%%%%%%%%%%%%%%%%%%%%%%%%%%%%
 %%%%%%%%%%%%%%%%%%%%%%%%%%%%%%%%%%%%%%%%%%%
 \section{Isospectral extension of GR} \label{sec:theory}
 %%%%%%%%%%%%%%%%%%%%%%%%%%%%%%%%%%%%%%%%%%%
 %%%%%%%%%%%%%%%%%%%%%%%%%%%%%%%%%%%%%%%%%%%
The analysis of \cite{Cano:2024wzo} singled out a particular EFT extension of GR with the property that GW propagation is non-birefringent in the large momentum regime. This theory contains a single quartic curvature term in the action, and it reads
\begin{align} \label{eq:isoeft}
	 S = \frac{1}{16\pi G}\int d^4 x \sqrt{|g|}\left[R + \alpha\left(\mathcal{C}^2+\tilde{\mathcal{C}}^2\right)\right]\, ,
\end{align}
where 
\begin{align}\label{eq:CCtilde}
	 	\mathcal{C}=R_{\mu \nu \rho \sigma} R^{\mu \nu \rho \sigma}, \quad \tilde{\mathcal{C}}=R_{\mu \nu \rho \sigma} \tilde{R}^{\mu \nu \rho \sigma}\,,
\end{align}
are quadratic invariants and
\begin{align}\label{dualRiem}
	 	\tilde{R}^{\mu \nu \rho \sigma}=\frac{1}{2} \epsilon^{\mu \nu \alpha \beta} R_{\alpha \beta}{ }^{\rho \sigma}\,,
 \end{align}
is the dual Riemann tensor. The coupling constant $\alpha$ has units of [length]$^6$ and we allow it to be arbitrary. Interestingly, it is possible to link this model to string theory, since, as noted in \cite{Cano:2024wzo}, the quartic correction in \req{eq:isoeft} is identical to the one arising from the four-dimensional compactification of type II superstring effective actions \cite{Gross:1986iv}.\footnote{In that context, $\alpha$ is fixed in terms of the string parameter $\alpha'$ according to $\alpha=\tfrac{\zeta(3)}{32}\alpha'^3$}
 
The equations of motion obtained from the variation of \req{eq:isoeft} are given by
\begin{align}\label{EFE}
\E_{\mu\nu}\equiv  G_{\mu\nu} +\alpha \mathcal{E}_{\mu\nu}^{(8)}=0\,,
\end{align}
where $G_{\mu\nu}$ is the Einstein tensor and the contribution from the higher-curvature terms reads
\begin{align}
 \mathcal{E}_{\mu\nu}^{(8)}=\tensor{P}{_{(\mu}^{\rho \sigma \gamma}} R_{\nu) \rho \sigma \gamma}-\frac{1}{2} g_{\mu \nu} \left(\mathcal{C}^2+\tilde{\mathcal{C}}^2\right)+2 \nabla^\sigma \nabla^\rho P_{(\mu|\sigma| \nu) \rho}\,,
\end{align}
where
\begin{align}
	& P_{\mu \nu \rho \sigma}\equiv \frac{\partial}{\partial R^{\mu\nu\rho\sigma}}\left(\mathcal{C}^2+\tilde{\mathcal{C}}^2\right)=4 \mathcal{C} R_{\mu \nu \rho \sigma}+2 \tilde{\mathcal{C}}\left(\tilde{R}_{\mu \nu \rho \sigma}+\tilde{R}_{\rho \sigma \mu \nu}\right)\, .
\end{align}
Since we are interested in perturbative corrections --- to first order in $\alpha$ --- to the solutions of the vacuum Einstein equations, we can simplify $\mathcal{E}_{\mu\nu}^{(8)}$ by assuming that it is evaluated on a Ricci-flat metric. For a Ricci-flat spacetime, the curvature satisfies the following identities
\begin{equation}\label{Ricciflatid}
\tilde{R}_{\mu \nu \rho \sigma}=\tilde{R}_{\rho \sigma \mu \nu}\, ,\quad \tensor{R}{_{\mu}^{\rho \sigma \gamma}} R_{\nu \rho \sigma \gamma}=\frac{1}{4}g_{\mu\nu}\mathcal{C}\, ,\quad \tensor{\tilde{R}}{_{\mu}^{\rho \sigma \gamma}} R_{\nu \rho \sigma \gamma}=\frac{1}{4}g_{\mu\nu}\tilde{\mathcal{C}}\, ,\quad \nabla^{\mu}R_{\mu\nu\alpha\beta}=0\, .
\end{equation}
Therefore, the correction to the equations can be simplified to
\begin{align}\label{Emunu8}
 \mathcal{E}_{\mu\nu}^{(8)}=+\frac{1}{2} g_{\mu \nu} \left(\mathcal{C}^2+\tilde{\mathcal{C}}^2\right)+8 R_{\mu\sigma\nu \rho}\nabla^\sigma \nabla^\rho \mathcal{C}+8 \tilde{R}_{\mu\sigma\nu \rho}\nabla^\sigma \nabla^\rho \tilde{\mathcal{C}} \,.
\end{align}
We next study the linearization of \req{EFE} around a generic curved background. 

\subsection{Effective equation for large momentum perturbations}
As shown in \cite{Cano:2024wzo}, the gravitational waves of large momentum in the theory \req{eq:isoeft} satisfy the dispersion relation
\begin{align}\label{disprel}
	k^2 = 64\alpha S_{\mu\nu}S^{\mu\nu}\, ,
\end{align}
where 
\begin{align} \label{eq:defofS1}
	S_{\mu\nu} &\equiv k^{\rho}k^{\sigma}R_{\mu\rho\nu\sigma}\,,
\end{align}
$k_{\mu}$ is the wave momentum, and $R_{\mu\rho\nu\sigma}$ is the background curvature tensor.  The crucial aspect about this theory is that the dispersion relation is independent of the polarization of the wave, and hence it is non-birefringent in the geometric optics limit. 

The central strategy of our paper is to consider an effective scalar equation that yields the dispersion relation \req{disprel} in the large momentum limit. Quite straighforwardly, we can write down the following equation
\begin{equation}\label{eq:wave_eq1}
\nabla^2\Phi+64\alpha\tensor{R}{^{\mu}_\alpha^\nu_\beta}\tensor{R}{^{\rho\alpha\sigma\beta}}\nabla_{\mu}\nabla_{\nu}\nabla_{\rho}\nabla_{\sigma}\Phi=0\, .
\end{equation}
The idea is that this equation describes gravitational perturbations of large momentum, and therefore we can solve it to find gravitational eikonal QNMs. In this section we put this approach on solid footing and provide a direct derivation of \req{eq:wave_eq1} from the linearized equations of motion.

Let us consider a perturbation $h_{\mu\nu}$ over a given background metric $g_{\mu\nu}$, 
\begin{equation}
g_{\mu\nu}^{\rm total}=g_{\mu\nu}+h_{\mu\nu}\, . 
\end{equation}
We consider the case in which the momentum of this perturbation is much larger than the curvature scale of the background.  Schematically, this implies that
\begin{equation}
|\nabla \nabla h|\gg |R h|\, ,
\end{equation}
where $R$ denotes the Riemann tensor and the indices can be placed arbitrarily. 
Equivalently, if the curvature scale is $R_{\mu\nu\alpha\beta}\sim 1/L^2$, this means that the wavelength of the perturbation is much shorter than $L$. 

We work in the  Lorentz gauge and additionally impose the metric perturbation to be traceless --- a condition that we will see to be consistent --- so that 
\begin{equation}\label{Lorentzgauge}
g^{\mu\nu}h_{\mu\nu}=0, \quad \nabla^{\mu}h_{\mu\nu}=0\, .
\end{equation}
In this gauge the linearized Einstein tensor reads
\begin{equation}\label{deltaGmunu}
\delta G_{\mu\nu}=-\frac{1}{2} \nabla^2 h_{\mu\nu}
-h^{\alpha\beta}R_{\alpha\mu\beta\nu}+h_{\alpha (\mu}R_{\nu)}^{\alpha}
+ \frac{1}{2} g_{\mu\nu} R_{\alpha\beta} h^{\alpha\beta}- \frac{1}{2} R\, h_{\mu\nu}\approx -\frac{1}{2}\nabla^2 h_{\mu\nu}\, ,
\end{equation}
where the ``$\approx$'' sign denotes that we kept only the leading terms for large momentum, and thus neglected the terms that depend explicitly on the background curvature. From now on, we will automatically discard the subleading terms in all expressions. 
Following this logic, in the linearization of $\mathcal{E}_{\mu\nu}^{(8)}$ in \req{Emunu8} we only keep the terms with the highest number of derivatives of $h_{\mu\nu}$, which in this case is four derivatives. Thus, we have\footnote{Observe that this tensor is symmetric on account of the first identity in \req{Ricciflatid}. }
\begin{equation}\label{eq:deltaE8v1}
\delta\mathcal{E}_{\mu\nu}^{(8)}= 8 R_{\mu\sigma\nu \rho}\nabla^\sigma \nabla^\rho \delta \mathcal{C}+8 \tilde{R}_{\mu\sigma\nu \rho}\nabla^\sigma \nabla^\rho \delta\tilde{\mathcal{C}}\, ,
\end{equation}
and taking into account that the linearized Riemann tensor reads

\begin{equation}
\delta R_{\alpha\beta\mu\nu}=-\nabla_{\alpha}\nabla_{[\mu}h_{\nu] \beta}-\nabla_{\beta}\nabla_{[\nu}h_{\mu] \alpha}\, ,
\end{equation}
we get
\begin{align}
\delta \mathcal{C}&=2R^{\alpha\beta\mu\nu}\delta R_{\alpha\beta\mu\nu}= 4 R^{\alpha\beta\mu\nu} \nabla_{\mu}\nabla_{\beta}h_{\nu\alpha}\, ,\\
\delta \tilde{\mathcal{C}}&=2\tilde{R}^{\alpha\beta\mu\nu}\delta R_{\alpha\beta\mu\nu}= 4 \tilde{R}^{\alpha\beta\mu\nu} \nabla_{\mu}\nabla_{\beta}h_{\nu\alpha}\, .
\end{align}
Therefore, keeping only the terms that contain four derivatives of $h_{\mu\nu}$, the linearized $\mathcal{E}_{\mu\nu}^{(8)}$ reads
\begin{equation}\label{eq:deltaE8v2}
\delta\mathcal{E}_{\mu\nu}^{(8)}=-32 \left[R_{(\mu|\sigma|\nu) \rho}R^{\alpha\lambda\beta\xi}+\tilde{R}_{(\mu|\sigma|\nu) \rho}\tilde{R}^{\alpha\lambda\beta\xi}\right]\nabla^{\sigma} \nabla^{\rho}\nabla_{\alpha}\nabla_{\beta} h_{\lambda\xi}\, ,
\end{equation}
where we added an explicit symmetrization in $\mu\nu$, which does not make a difference in the large momentum limit but makes $\delta\mathcal{E}_{\mu\nu}^{(8)}$ manifestly symmetric. 
The product $\tilde{R}\tilde{R}$ can then be expanded using the properties of the Levi-Civita tensor
\begin{equation}
\tilde{R}_{\mu\sigma\nu \rho}\tilde{R}^{\alpha\lambda\beta\xi}=\frac{1}{4}\epsilon_{\mu\sigma\tau\epsilon}\epsilon^{\alpha\lambda\gamma\pi}\tensor{R}{^{\tau\epsilon}_{\nu \rho}}\tensor{R}{_{\gamma\pi}^{\beta\xi}}=-\frac{1}{4}\delta^{\alpha\lambda\gamma\pi}_{\mu\sigma\tau\epsilon}\tensor{R}{^{\tau\epsilon}_{\nu \rho}}\tensor{R}{_{\gamma\pi}^{\beta\xi}}=-3!\delta^{[\alpha}_{\mu}\delta^{\lambda}_{\sigma}\tensor{R}{^{\gamma\pi]}_{\nu \rho}}\tensor{R}{_{\gamma\pi}^{\beta\xi}}\, .
\end{equation}
We can then expand the antisymmetrization and simplify the result by using the properties satisfied by the Riemann tensor and the metric perturbation. Since we are evaluating the higher-derivative part of the equations and we are only interested in the first-order corrections, we can assume that both the curvature and the metric perturbation are given by the GR values. This implies that we can assume $R_{\mu\nu}=0$ and $\nabla^2 h_{\mu\nu}=0$ in addition to the gauge conditions \req{Lorentzgauge}. We can also commute covariant derivatives, since the commutator yields a term that is subleading in the large momentum limit. Using all of these properties, $\delta\mathcal{E}_{\mu\nu}^{(8)}$ is reduced to the sum of four terms, and the total linerized equations (including the Einstein term) read
\begin{equation}
\begin{aligned}
\delta\mathcal{E}_{\mu\nu}%&=-\frac{1}{2}\nabla^2 h_{\mu\nu} -16\gamma (R_{\mu\sigma\nu \rho}R^{\alpha\lambda\beta\xi}+\tilde{R}_{\mu\sigma\nu \rho}\tilde{R}^{\alpha\lambda\beta\xi})\nabla^{\sigma} \nabla^{\rho}\nabla_{\alpha}\nabla_{\beta} h_{\lambda\xi}\\
&=-\frac{1}{2}\nabla^2 h_{\mu\nu} -32\alpha \Big[R_{(\mu|\sigma|\nu) \rho}R^{\alpha\lambda\beta\xi}\nabla^{\sigma} \nabla^{\rho}\nabla_{\alpha}\nabla_{\beta} h_{\lambda\xi}+\tensor{R}{_{\gamma\pi}^{\beta\xi}} \tensor{R}{_{(\nu| \rho}^{\lambda\pi}}\nabla^{\gamma} \nabla^{\rho}\nabla_{|\mu)}\nabla_{\beta} h_{\lambda\xi}
\\
&+\tensor{R}{_{\gamma\pi}^{\beta\xi}} \tensor{R}{_{(\nu| \rho}^{\pi\alpha}}\nabla^{\gamma} \nabla^{\rho}\nabla_{\alpha}\nabla_{\beta} h_{|\mu)\xi}
+\tensor{R}{_{\gamma(\mu}^{\beta\xi}} \tensor{R}{_{\nu) \rho}^{\alpha\lambda}}\nabla^{\gamma} \nabla^{\rho}\nabla_{\alpha}\nabla_{\beta} h_{\lambda\xi}\Big]\, .
\end{aligned}
\end{equation}
In order to make further progress, it is convenient now to go to momentum space so that $\nabla_{\mu}\to i k_{\mu}$. Straightforwardly, we get
\begin{equation}\label{deltaE3}
\begin{aligned}
\delta\mathcal{E}_{\mu\nu}&=\frac{1}{2}k^2 h_{\mu\nu} -32\alpha \Big(S_{\mu\nu}S^{\alpha\beta} h_{\alpha\beta}+\tensor{S}{_{\alpha}^{\beta}} \tensor{R}{_{(\nu| \rho}^{\lambda\alpha}} k^{\rho}k_{|\mu)} h_{\lambda\beta}+\tensor{S}{_{\lambda}^{\beta}} \tensor{S}{_{(\nu}^{\lambda}} h_{\mu)\beta}
-\tensor{S}{_{\mu}^{\alpha}} \tensor{S}{_{\nu}^{\beta}}  h_{\alpha\beta}\Big)\, ,
\end{aligned}
\end{equation}
where we recall that $S_{\mu\nu}$ is defined in \req{eq:defofS1}. 
Observing that $g^{\mu\nu}\delta\mathcal{E}_{\mu\nu}=0$ --- which follows from the tracelessness of $S_{\mu\nu}$ --- we conclude that the trace-free condition \eqref{Lorentzgauge} for $h_{\mu\nu}$ is consistent. 
These equations \eqref{deltaE3} can be simplified if we focus on the components orthogonal to $k_{\mu}$. In GR, the transverse condition \req{Lorentzgauge} still allows for $h_{\mu\nu}$ to have components in the direction of $k_{\mu}$, since $k_{\mu}$ is null and hence transverse to itself. In order to remove those components and retain only the physical degrees of freedom, we project the equations on the directions transverse to $k_{\mu}$. To this end, let $\tensor{P}{_{\mu\nu}^{\alpha\beta}}$ be a projector satisfying
\begin{equation}
\tensor{P}{_{\mu\nu}^{\alpha\beta}}k_{\beta}=\tensor{P}{_{\mu\nu}^{\alpha\beta}}k^{\mu}=g^{\mu\nu}\tensor{P}{_{\mu\nu}^{\alpha\beta}}=g_{\alpha\beta}\tensor{P}{_{\mu\nu}^{\alpha\beta}}=0\, .
\end{equation}
We define 
\begin{equation}
\hat{h}_{\mu\nu}=\tensor{P}{_{\mu\nu}^{\alpha\beta}}h_{\alpha\beta}\, ,
\end{equation}
which contains the physical degrees of freedom of the metric perturbation --- namely the two polarization modes of the graviton. The projection of the equations \req{deltaE3} yields 
\begin{equation}\label{deltaE4}
\begin{aligned}
\tensor{P}{_{\mu\nu}^{\mu'\nu'}}\delta\mathcal{E}_{\mu'\nu'}&=\frac{1}{2}k^2 \hat{h}_{\mu\nu} -32\alpha \tensor{P}{_{\mu\nu}^{\mu'\nu'}}\Big(S_{\mu'\nu'}S^{\alpha\beta} h_{\alpha\beta}+\tensor{S}{_{\lambda}^{\beta}} \tensor{S}{_{\nu'}^{\lambda}} h_{\mu'\beta}
-\tensor{S}{_{\mu'}^{\alpha}} \tensor{S}{_{\nu'}^{\beta}}  h_{\alpha\beta}\Big)\, ,
\end{aligned}
\end{equation}
and this result can be further simplified by using the following antisymmetrization identities
\begin{align}\notag	
	0=k^{\rho}k_{[\sigma|}\tensor{S}{_{\mu'}^{\nu'}}\tensor{R}{_{|\alpha \rho|}^{ \sigma \beta}}\tensor{P}{_{\mu\nu}^{\mu'}_{|\nu'}}\tensor{h}{^{\alpha}_{\beta]}}=\frac{1}{30}\tensor{P}{_{\mu\nu}^{\mu'\nu'}}\Big[&- S_{\alpha\beta}h^{\alpha\beta}S_{\mu'\nu'} + S^{\alpha}{}_{\mu'}S^{\beta}{}_{\nu'}h_{\alpha\beta}\\
	&+ S_{\nu'\lambda}S^{\beta\lambda} h_{\mu'\beta}\Big] \, ,
	\\\notag
	0=k^{\rho}k_{[\sigma}\tensor{S}{_{\mu'|}^{\nu'}}\tensor{R}{_{|\alpha| \rho}^{ \sigma \beta}}\tensor{P}{_{\mu\nu}^{\mu'}_{|\nu'}}\tensor{h}{^{\alpha}_{\beta]}}=\frac{1}{60}\tensor{P}{_{\mu\nu}^{\mu'\nu'}}\Big[&-2 S_{\alpha\beta}h^{\alpha\beta}S_{\mu'\nu'} + 2 S^{\alpha}{}_{\mu'}S^{\beta}{}_{\nu'}h_{\alpha\beta}\\
	&+ 4 S_{\nu'\lambda}S^{\beta\lambda} h_{\mu'\beta}-S_{\alpha\beta}S^{\alpha\beta}h_{\mu'\nu'}\Big] \, .
\end{align}
To derive these results we have taken into account the relations $k^2=0$, $k^{\mu}S_{\mu\nu}=0$, $g^{\mu\nu}S_{\mu\nu}=0$, which hold at zeroth order in $\alpha$. 
Using these identities in \req{deltaE4}, we get
\begin{equation}\label{deltaEfinal}
\begin{aligned}
\tensor{P}{_{\mu\nu}^{\mu'\nu'}}\delta\mathcal{E}_{\mu'\nu'}&=\frac{1}{2}k^2 \hat{h}_{\mu\nu} -32\alpha S_{\alpha\beta}S^{\alpha\beta} \hat{h}_{\mu\nu}=0\, .
\end{aligned}
\end{equation}
From this equation, the dispersion relation \req{disprel} follows straightforwardly, and we obtain additional information;  Replacing $k_{\mu}$ back by the covariant derivative, \req{deltaEfinal} implies that the transverse and traceless part of $h_{\mu\nu}$ satisfies the equation 
\begin{equation}\label{hhateq}
\left(\nabla^2+64\alpha \tensor{R}{^{(\lambda}_\alpha^\eta_\beta}\tensor{R}{^{\rho|\alpha|\sigma)\beta}}\nabla_{\lambda}\nabla_{\eta}\nabla_{\rho}\nabla_{\sigma}\right)\hat{h}_{\mu\nu}=0\, .
\end{equation}
The key observation here is that the differential operator does not act explicitly on the indices of $\hat{h}_{\mu\nu}$, and this is the reason behind the isospectrality of this theory in the eikonal limit; both polarization modes satisfy the same equation.  Furthermore, in the large momentum limit, the effect of the indices of $\hat{h}_{\mu\nu}$ is irrelevant, so in practice we can replace \req{hhateq} by an effective scalar equation
\begin{equation}\label{effectivePhieq}
\left(\nabla^2+64\alpha \tensor{R}{^{(\lambda}_{\alpha}^{\eta}_{\beta}}\tensor{R}{^{\rho|\alpha|\sigma)\beta}}\nabla_{\lambda}\nabla_{\eta}\nabla_{\rho}\nabla_{\sigma}\right)\Phi=0\, .
\end{equation}

\subsection{Rotating black holes}
The equation \req{effectivePhieq} holds for perturbations around all vacuum solutions of \req{eq:isoeft}, but we are interested in the case of rotating black hole solutions. The Kerr metric is no longer a solution of \req{eq:isoeft}, and so we need to obtain the corrections to the Kerr metric \cite{Cano:2019ore} in order to evaluate \req{effectivePhieq}. At first order in $\alpha$, the corrections to the background geometry will affect the form of the $\nabla^2$ operator \cite{Cano:2020cao}. However, in the large momentum limit, the correction to $\nabla^2$ is subleading compared with the $\nabla\nabla\nabla\nabla$ term in \req{effectivePhieq}, since the latter scales with the fourth power of the momentum, and the former with the square. 

Therefore, at leading order in large momentum, we can disregard the corrections to the background metric and solve \req{effectivePhieq} on the Kerr background. We recall that the Kerr metric takes the form
\begin{align}\label{eq:Kerr}
	ds^2 = -\frac{\Delta}{\Sigma}\left(dt-a (1-x^2)d\phi\right)^2 + \frac{\Sigma}{\Delta}dr^2 + \frac{\Sigma}{1-x^2}dx^2 + \frac{1-x^2}{\Sigma}\left(adt-(r^2+a^2)d\phi\right)^2\, ,
\end{align}
where $x=\cos \theta$ and 
\begin{align} \label{eq:DeltaWsol}
	\Delta = r^2 - 2 M r + a^2, \qquad \Sigma = r^2 + a^2 x^2\, .
\end{align}
Here $M$ is the mass of the black hole and $a M$ is its angular momentum. We will also often use the dimensionless spin parameter
\begin{equation}
\chi=\frac{a}{M}\, ,
\end{equation}
that takes values from $0$ (no rotation) to $1$ (extremality).  We also introduce the dimensionless coupling constant
\begin{equation}\label{hatalpha}
\hat{\alpha}=\frac{\alpha}{M^6}\, ,
\end{equation}
and we restrict to the regime in which $|\hat\alpha|\ll 1$. 

%%%%%%%%%%%%%%%%%%%%%%%%%%%%%%%%%%%%%%%%%%%%%%%%
%%%%%%%%%%%%%%%%%%%%%%%%%%%%%%%%%%%%%%%%%%%%%%%%
\section{Eikonal QNMs from geometric optics}\label{sec:geometricoptics}
%%%%%%%%%%%%%%%%%%%%%%%%%%%%%%%%%%%%%%%%%%%%%%%%%
%%%%%%%%%%%%%%%%%%%%%%%%%%%%%%%%%%%%%%%%%%%%%%%
It is a well-established fact that eikonal quasinormal modes of Kerr black holes can be linked to properties of null geodesics in the photon sphere \cite{Yang:2012he}.  This correspondence has its origin on the fact that gravitational waves of large momentum move along null geodesics in GR, which in turn is a consequence of the wave-like form of the linearized Einstein equations, as reflected in \req{deltaGmunu}. However, the GW propagation is modified in the theory \req{eq:isoeft}, as the momentum is no longer null according to the dispersion relation \req{disprel}. It also follows that the momentum is not geodesic either, $k^{\mu}\nabla_{\mu}k^{\alpha}\neq 0$.  This implies, in particular, that the set of unstable closed GW orbits around the black hole is now different to the usual photon sphere. We may call this set of unstable GW orbits the \emph{graviton-sphere} \cite{Papallo:2015rna,Chowdhury:2024uzd}, in order to distinguish it from the former. 
We expect that there is a correspondence between the properties of the graviton-sphere orbits and the quasinormal modes in the eikonal limit. In this section, we formalize this intuition and compute QNMs from the graviton-sphere orbits of rotating black holes in the theory \req{eq:isoeft}.

In order to study the GW trajectories around black holes, we will consider the geometric optics limit of the master equation \req{effectivePhieq}. 
To this end, we consider the ansatz
\begin{align}\label{geomphi}
	\Phi=A e^{i S},
\end{align}
and we consider the regime in which the phase $S$ varies in a much shorter length scale than the amplitude $A$. In this regime, the gradient of $S$ is identified as the momentum vector of a bundle of GW orbits,
\begin{equation}\label{kmuS}
k_{\mu}=\partial_{\mu}S\, ,
\end{equation}
and individual GW trajectories can be obtained by finding the integral curves of $k_{\mu}$, 
\begin{equation} \label{eq:kSrelation}
	\frac{dx^{\mu}}{d\lambda}=k^{\mu}=g^{\mu\nu}\partial_{\nu}S\, ,
\end{equation}
where $\lambda$ is an affine parameter. Since $A$ is an slowly-varying function we have $|\partial_{\mu} A|\ll |k_{\mu}|$, and likewise we demand that the rate of change of $k_{\mu}$ is much smaller than $k_{\mu}$ itself, schematically $|\nabla k|\ll |k k|$.  We can then plug \req{geomphi} into \req{effectivePhieq} and expand it in the geometric optics limit.  At leading order --- disregarding all the derivatives of $A$ and $k_{\mu}$ --- we get precisely the dispersion relation for $k_{\mu}$, 
\begin{align} \label{eq:dispersion}
	g^{\mu\nu}k_{\mu}k_{\nu} - 64 \alpha k^{\mu}k^{\nu}k_{\rho}k_{\sigma}R_{\gamma \mu \tau \nu}R^{\gamma \rho \tau \sigma}=0\, .
\end{align}
At next-to-leading order, we get an equation for the rate of change of the amplitude,
\begin{align} \label{eq:dispersionsimp}
	\nabla_\mu k^\mu + 2 k^\mu \nabla_\mu \log A &= 
	64\alpha R^{(\mu}{}_{\alpha}{}^{\nu}{}_{\beta}R^{\rho|\alpha|\sigma)\beta} \left(6  k_\mu k_\nu \nabla_\rho k_\sigma + 4  k_\mu k_\nu k_\rho \nabla_\sigma \log A\right)\,.
\end{align}
These two equations determine the real and imaginary parts of the QNM frequencies, as we discuss next. 

\subsection{Leading order: real part of the frequency}
When we take into account \req{kmuS}, the dispersion relation as given in \req{eq:dispersion} becomes in fact the Hamilton-Jacobi equation for $S$. In the Kerr background in GR ($\alpha=0$), this equation is separable and one can find a full analytic solution, which leads to a complete set of first integrals of the geodesic motion.  When the corrections are included, the Hamilton-Jacobi equation is in general not separable, except if we restrict to motion in the equatorial plane, \textit{i.e.}, $x=0$. Thus, we will focus on this case for simplicity. Since $t$ and $\phi$ are isometric coordinates, we can automatically separate them and consider the following ansatz in the equatorial plane
\begin{align} \label{eq:SansatzforKerr}
	S = - E t + S_r (r) + L_z \phi\, ,
\end{align}
where $E$ and $L_{z}$ are constants representing intuitively the energy and angular momentum of the orbit. From the point of view of the field $\Phi$ given in \req{geomphi}, it is clear that these quantities are the frequency and the angular harmonic number $m$,  
\begin{equation}\label{ELzrel}
E=\omega_{R}\, ,\qquad L_{z}=m\, .
\end{equation}
Plugging \req{eq:SansatzforKerr} into the Hamilton-Jacobi equation \req{eq:dispersion} and evaluating at $x=0$, we get an equation for the radial function $S_{r}$. The contribution of the GR part reads
\begin{align} \label{eq:ksquaredexpanded}
	g^{\mu\nu}k_{\mu}k_{\nu} = \frac{\Delta}{r^2}(S'_r)^2-\frac{\left(E(r^2+a^2)-a L_{z}\right)^2 }{\Delta  r^2}+\frac{\left(L_{z}-a E\right)^2}{r^2}\,.
\end{align}
which has been evaluated on the uncorrected Kerr background \req{eq:Kerr}. The prime in $S'_{r}$ denotes a derivative with respect to $r$. Therefore, in the case of GR, where the dispersion relation is simply $k^2=0$, we get
\begin{equation}
\left(S_{r}^{\rm GR}{}'\right)^2=\frac{\left( E(r^2+a^2)-a L_{z}\right)^2 }{\Delta^2}-\frac{\left(L_{z}-a E\right)^2}{\Delta}\,.
\end{equation}
Since we are only interested in first-order-in-$\alpha$ corrections, we can evaluate the higher-derivative terms in \eqref{eq:dispersion} on this value of $S_{r}$. We get the simple expression
\begin{align} \label{eq:deltaVorrection}
	k^{\mu}k^{\nu}k_{\rho}k_{\sigma}R_{\gamma \mu \tau \nu}R^{\gamma \rho \tau \sigma} = \frac{18M^2 \left(L_{z}-a E\right){}^4}{r^{10}}.
\end{align}
Putting everything together, the full dispersion relation \eqref{eq:dispersion} yields
\begin{equation}
\frac{\Delta}{r^2}(S'_r)^2-\frac{\left(E(r^2+a^2)-a L_{z}\right)^2 }{\Delta  r^2}+\frac{\left(L_{z}-a E\right)^2}{r^2}-\alpha \frac{1152M^2 \left(L_{z}-a E\right){}^4}{r^{10}}+\mathcal{O}(\alpha^2)=0\, ,
\end{equation}
which straightforwardly determines $S_{r}(r)$. Now, taking into account that the GW trajectories are the solutions of \req{eq:kSrelation}, we have the following equation for the radial motion,  
\begin{equation}
\frac{dr}{d\lambda}=k^{r}=\frac{\Delta}{r^2}S'_{r}\,,
\end{equation} 
which can more interestingly be written in the form 
\begin{align} \label{eq:rdotsquaredplusV}
\left(\frac{dr}{d\lambda}\right)^2 +  \mathcal{U} = 0\, ,
\end{align}
where we have an effective potential
\begin{align} \label{eq:potsol}
 \mathcal{U}%&=-\left(\frac{\Delta}{r^2}\dot{S}_r\right)^2%
 =-\frac{\left( E(r^2+a^2)-a L_{z}\right)^2 }{r^4}+\frac{\Delta \left(L_{z}-a E\right)^2}{r^4}-\alpha \frac{1152M^2 \Delta \left(L_{z}-a E\right){}^4}{r^{12}}\, .
\end{align}
In this form, we then search for circular orbits by imposing the conditions
\begin{align} \label{eq:VandVprime}
	\mathcal{U} (r_0) = 0, \qquad \mathcal{U}'(r_0) = 0,
\end{align}
where $r_0$ is the radius of the orbit to be determined. In order to solve these equations, we assume an expansion in $\alpha$ of both the energy of the orbit and the radius, 
\begin{align} \label{eq:omegar0expansion}
	E = E^{\rm Kerr} + \alpha \delta E, \qquad r_0 = r_{0}^{\rm Kerr} + \alpha \delta r_0.
\end{align}
At zeroth order in $\alpha$ (the GR case), we find that the radius of the orbit is fixed by the equation 
\begin{equation}
\chi=\frac{\sqrt{\rho}}{2}(3-\rho)\, , \quad \text{where} \quad \rho=\frac{ r_{0}^{\rm Kerr} }{M}\, ,
\end{equation}
whose solution can be expressed analytically as
\begin{equation}\label{rhosol}
\rho=2\left[1+\cos\left(\frac{2}{3}\arccos(-\chi)\right)\right]\, .
\end{equation}
On the other hand, the energy of the orbit reads
\begin{align} \label{eq:omegaRKerrsol}
	E^{\rm Kerr} = \frac{2 L_{z}}{M\sqrt{\rho}(3+\rho)}\, .
\end{align}
At first order in $\alpha$, we end up with two linearized equations
\begin{align}
	0 &= -\frac{4608 L_z^4 (1-\rho )^2}{M^8\rho ^7 (3+\rho )^4} + \frac{L_{z} (1-\rho ) \delta E}{M\sqrt{\rho ^3}},
	\\
	0&=\frac{18432 L_{z}^4 (1-\rho ) (5-3 \rho )}{M^9\rho ^8 (3+\rho )^4}-\frac{24 \delta r_0 L_{z}^2 }{M^4\rho ^3 (3+\rho )^2}+\frac{2 L_{z} (\rho -3) \delta E}{M^2\sqrt{\rho ^5}},
\end{align}
that we must solve for to obtain $\delta E$ and $\delta r_0$. The corrections to the energy  and radius are therefore given by
\begin{align}\label{dEsol}
	\delta E &= \frac{4608 L_{z}^3 (1-\rho )}{M^7\rho ^{11/2} (3+\rho )^4}\,, \\\label{drsol}
	 \delta r_0 &= \frac{384 L_{z}^2 (7 -5 \rho ) (1-\rho )}{M^5\rho ^5 (3+\rho )^2}.
\end{align}
According to \req{ELzrel}, $\delta E$ correspond to the correction to the real part of the QNM frequency. 

\subsection{Next-to-leading order: imaginary part of the frequency}

We now turn to the imaginary part of the frequency, which should be related to the instability timescale of the circular orbits we just studied. To understand the decay of a bundle of GWs following those orbits, we must consider the next-to-leading order corrections in the geometric optics limit, \textit{i.e.}, the equation \req{eq:dispersionsimp} that rules the evolution of the amplitude. In order to evaluate that equation, we note that
\begin{equation}
k^{\mu}=\left(\dot{t},\, \dot{r},\, 0, \, \dot{\phi}\right)\, ,
\end{equation}
with 
\begin{align}\label{tdot}
\dot{t}&=\frac{(r^2+a^2)\left(E(r^2+a^2)-a L_{z}\right)}{r^2\Delta}+\frac{a(L_{z}-a E)}{r^2}\, ,\\\label{rdot}
\dot{r}&=\pm \sqrt{-\mathcal{U}}\, ,\\\label{pdot}
\dot{\phi}&=\frac{L_{z}-a E}{r^2}+\frac{a\left(E(r^2+a^2)-a L_{z}\right)}{r^2\Delta}\, .
\end{align}
Since we are restricting to motion in the equatorial plane, a sufficient and separable ansatz for $A$ near $r=r_0$ is 
\begin{align} \label{eq:Aansatz}
	A = e^{-\gamma t}(r-r_0)^n\, ,
\end{align}
where $\gamma$ is a constant that determines the damping time. The radial dependence $(r-r_{0})^n$ simply represents the first term in the Taylor series of $A$ around $r_0$, and it is characterized by an integer number $n=0,1,2,\ldots$. In the QNM correspondence, this integer labels the overtone index of the QNM, while $\gamma$ is clearly minus the imaginary part of the frequency, 
\begin{equation}\label{gammaomega}
\gamma=-\omega_{I}\, .
\end{equation}
 Using these expressions, the left hand side of \eqref{eq:dispersionsimp} can then be recast as
\begin{align} \label{eq:correction3}
	\nabla_\mu k^\mu + 2 k^\mu \nabla_\mu \log A = -\frac{1}{2}\frac{\mathcal{U} '}{\sqrt{-\mathcal{U} }} -  \frac{2}{r}\sqrt{-\mathcal{U} } + 2\left(-\dot{t}\gamma + \dot{r}n(r-r_0)^{-1}\right)\, .
\end{align}
Since in \req{eq:Aansatz} we are considering only the leading term in the Taylor expansion of $A$, the equation \eqref{eq:correction3} must be evaluated at the orbital radius $r=r_0$. To this end, we make use of \eqref{eq:VandVprime} and expand the potential via
\begin{align}
	\mathcal{U} (r) &\simeq \frac{1}{2}\mathcal{U}''(r_0)(r-r_0)^2,
	\qquad
	\mathcal{U} '(r) \simeq \mathcal{U}''(r_0)(r-r_0)\,.
\end{align} 
After simplifying, we find
\begin{align} \label{eq:auxresult1}
	\nabla_\mu k^\mu + 2 k^\mu \nabla_\mu \log A = - 2\dot{t}\gamma + 2\left(n+\frac{1}{2}\right)\sqrt{-\mathcal{U}''(r_0)/2}\,.
\end{align}
Although the right hand side of \eqref{eq:dispersionsimp} is more nontrivial, we only need to evaluate it on the Kerr background. The correction in \eqref{eq:dispersionsimp} containing $\log A$ gives us
\begin{align} \label{eq:correction1}
	4R^{(\mu}{}_{\alpha}{}^{\nu}{}_{\beta}R^{\rho|\alpha|\sigma)\beta}  k_\mu k_\nu k_\rho \nabla_\sigma \log A &= \frac{36M^2 (L_{z}-a E)^2}{r^8}\left[\left(-\dot{t} +
	\frac{2a}{r^{2}}(L_{z}-a E)\right)\gamma  + \frac{\dot{r} n}{(r-r_0)}\right]\,.
\end{align}
The second correction on the right hand side of \eqref{eq:dispersionsimp} involves covariant derivatives acting on $k^{\mu}$ and takes the form
\begin{align} \label{eq:correction2}
	\begin{split}
		6R^{(\mu}{}_{\alpha}{}^{\nu}{}_{\beta}R^{\rho|\alpha|\sigma)\beta}  k_\mu k_\nu \nabla_\rho k_\sigma &= \frac{9 M^2 (L_{z}-a E)^2}{r^8}\frac{\mathcal{U}'}{\sqrt{-\mathcal{U}}}
		\\&+\frac{60 M^2 \left(a (L_{z}-a E)-r^2 E \right)^2}{r^9 \Delta}\sqrt{-\mathcal{U}}-\frac{60 M^2 (\sqrt{-\mathcal{U}})^{3}}{r^5 \Delta}\,.
	\end{split}
\end{align}
Evaluating once again at $r=r_0$, we find only the first term on the right hand side contributes at leading order
\begin{align} \label{eq:auxresult2}
	6R^{(\mu}{}_{\alpha}{}^{\nu}{}_{\beta}R^{\rho|\alpha|\sigma)\beta}  k_\mu k_\nu \nabla_\rho k_\sigma &=\frac{9 M^2 (L_{z}-a E )^2}{r^8}\sqrt{-\mathcal{U}''(r_0)/2} \, .
\end{align}
Combining the expressions \eqref{eq:auxresult1}, \req{eq:correction1} (evaluated at $r=r_0$) and \eqref{eq:auxresult2}, we find a relatively simple equation
\begin{align}
	\left(n+\frac{1}{2}\right)\sqrt{-\mathcal{U}''/2}\left(1+128\alpha p \right) = \left(\left(1+ 128\alpha p\right)\dot{t}-2304\alpha\frac{aM^2(L_{z}-a E)^3}{r_0^{10}}\right)\gamma\,,
\end{align}
where $p=9M^2(L_{z}-a E)^2/r_0^8$. We note that the factor $(1+128\alpha p)$ appears on both sides of the equation and therefore can be dropped at first order in $\alpha$, leading to the following result for $\gamma$:
\begin{align}\label{gammasol}
	\gamma = \left(n+\frac{1}{2}\right)\frac{\sqrt{-\mathcal{U}''(r_0)/2}}{\dot{t}-2304\alpha aM^2 (L_{z}-a E)^3/r_0^{10}}\, .
\end{align}
Interestingly, due to the correction to $\dot{t}$ in the denominator, this quantity is no longer the Lyapunov exponent that one would associate to the orbits described by \req{tdot}, \req{rdot} and \req{pdot}. In fact, the Lyapunov exponent for these circular orbits is \cite{Cardoso:2008bp,Yang:2012he}
\begin{equation}
\gamma_{L}=\dot{t}^{-1}\sqrt{-\mathcal{U}''(r_0)/2}\, ,
\end{equation} 
and so we see that $\gamma\neq (n+1/2)\gamma_{L}$. Therefore, the connection between the damping time of a bundle of GWs and the usual geometric Lyapunov exponent is broken. 

Our final observation is that the denominator in \req{gammasol} is proportional to the derivative of the potential with respect to the frequency. In fact, we observe
\begin{align}
	\dot{t}= -\frac{r^2}{2\Delta}\partial_{E}\mathcal{U}(r_0) + 2304\alpha\frac{aM^2(L_{z}-a E)^3}{r_0^{10}}\, ,
\end{align}
and therefore we can write $\gamma$ in the compact form
\begin{align}\label{gammamaster}
	\gamma = -\left(n+\frac{1}{2}\right)\left.\frac{\Delta\sqrt{-2\mathcal{U}''}}{r^2 \partial_{E} \mathcal{U}}\right|_{r=r_0}.
\end{align}
Inserting in this formula the expressions for $E$ and $r_0$ obtained in the previous section --- see  \req{eq:omegar0expansion}, \eqref{rhosol}, \eqref{eq:omegaRKerrsol}, \req{dEsol}, \req{drsol} --- we get, at first order in $\alpha$
\begin{align}
	\gamma= \left(n+\frac{1}{2}\right)\left[\frac{\sqrt{3} (\rho -1)}{M \rho  (3 +\rho )}+\alpha\frac{192 \sqrt{3} L_{z}^2 (1-\rho )^2 \left( 35 \rho ^2 + 22 \rho - 144\right)}{M^7\rho ^7 (3 +\rho )^4}+\mathcal{O}(\alpha^2)\right]\, .
\end{align}

\subsection{Summary: QNM frequencies with $\ell=m$}
Taking into account \req{ELzrel} and \req{gammaomega}, $E$ and $-\gamma$ are respectively the real and imaginary parts of QNM frequencies with $\ell=m=L_{z}$ in the eikonal limit --- the fact that $\ell=m$ follows from the corresponding orbits being equatorial. 
For convenience, we collect our results here. At first order in the corrections, the real and imaginary parts of the frequencies read
\begin{equation}
\omega_{R}=\omega_{R}^{\rm Kerr}+\hat{\alpha}\delta\omega_{R}\, , \quad  \omega_{I}=\omega_{I}^{\rm Kerr}+\hat{\alpha}\delta\omega_{I}\, ,
\end{equation}
where we recall that $\hat\alpha=\alpha/M^6$ and 
\begin{align}\label{omegaRgeom}
	\omega_{R}^{\rm Kerr}&=\frac{2 \ell }{M\sqrt{\rho}(3+\rho)}\, , \quad &\delta \omega_R &= \frac{4608 \ell^3 (1-\rho )}{M\rho ^{11/2} (3+\rho )^4}\, ,\\\label{omegaImgeom}
	\omega_{I}^{\rm Kerr}&=-\left(n+\frac{1}{2}\right)\frac{\sqrt{3} (\rho -1)}{M \rho  (3 +\rho )}\, ,\quad &\delta\omega_{I}&=-\left(n+\frac{1}{2}\right)\frac{192\sqrt{3} \ell^2 (1-\rho )^2 \left( 35 \rho ^2 + 22 \rho - 144\right)}{M\rho ^7 (3 +\rho )^4}\,.
\end{align}

\begin{figure}[h!] 
	\includegraphics[width=8cm]{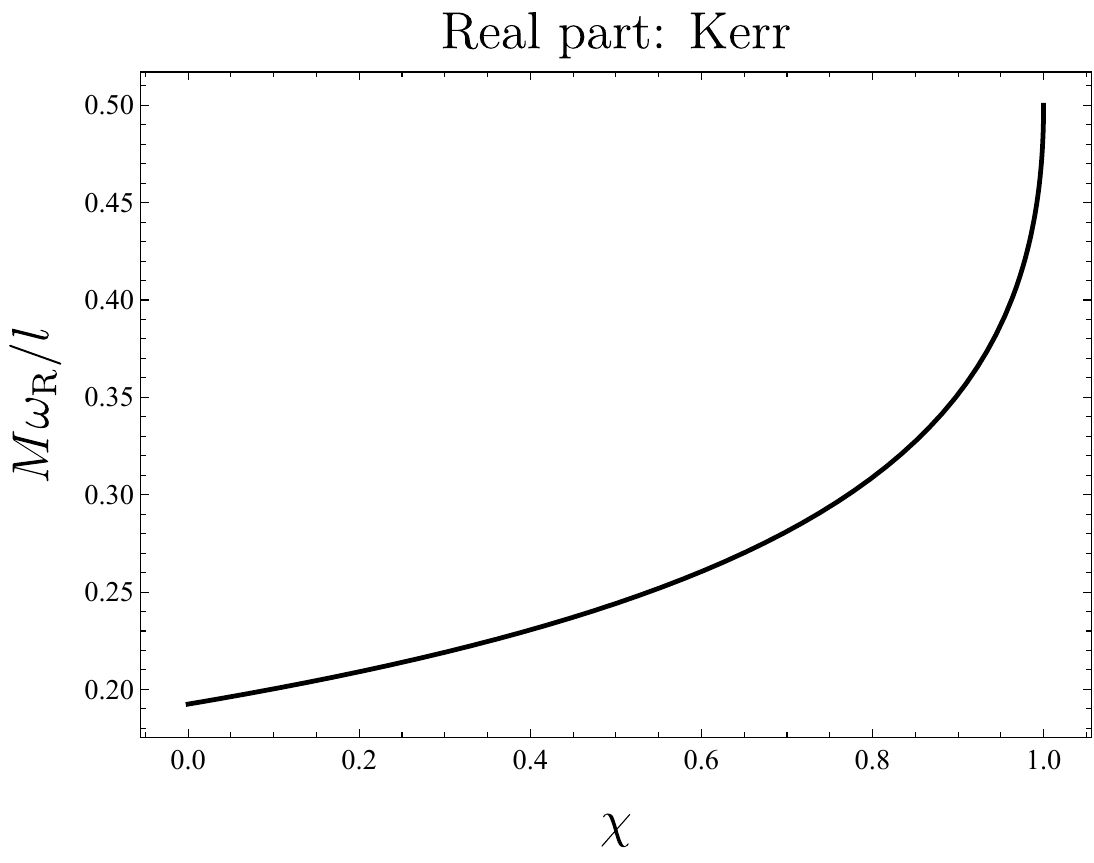}
	\includegraphics[width=8cm]{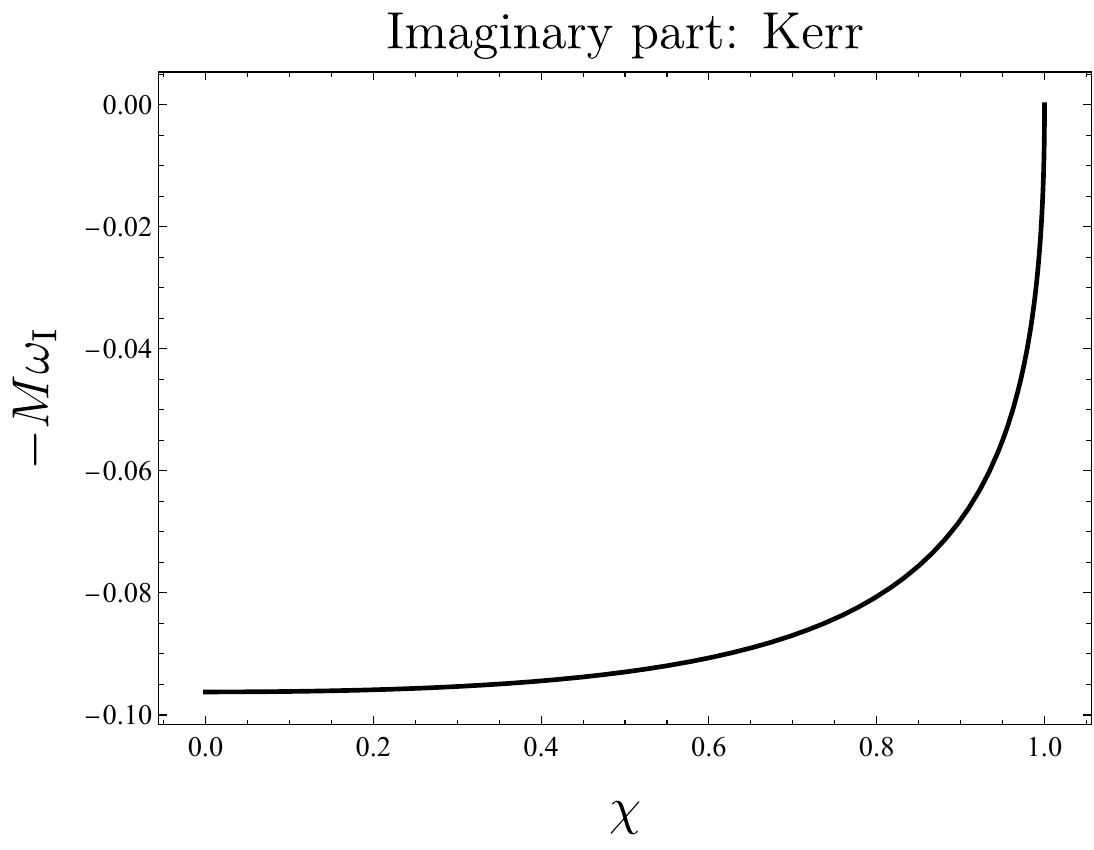}
	\includegraphics[width=8cm]{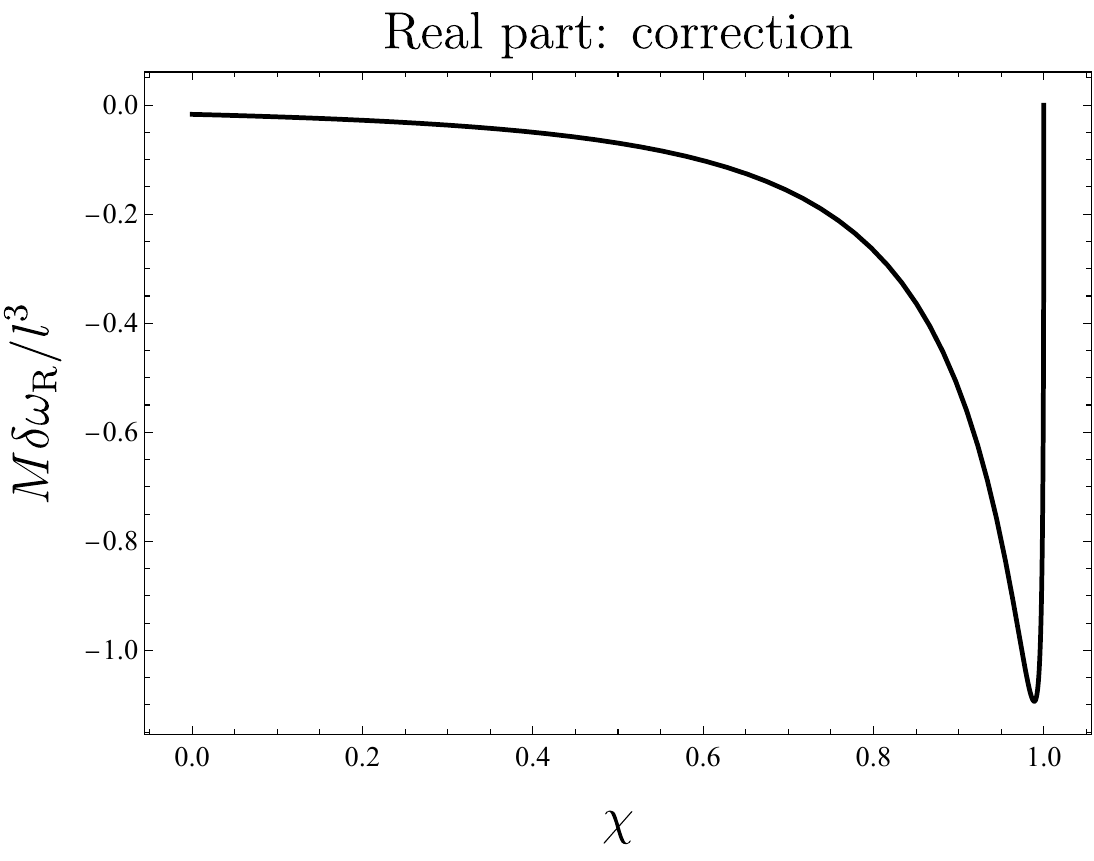}
	\includegraphics[width=8cm]{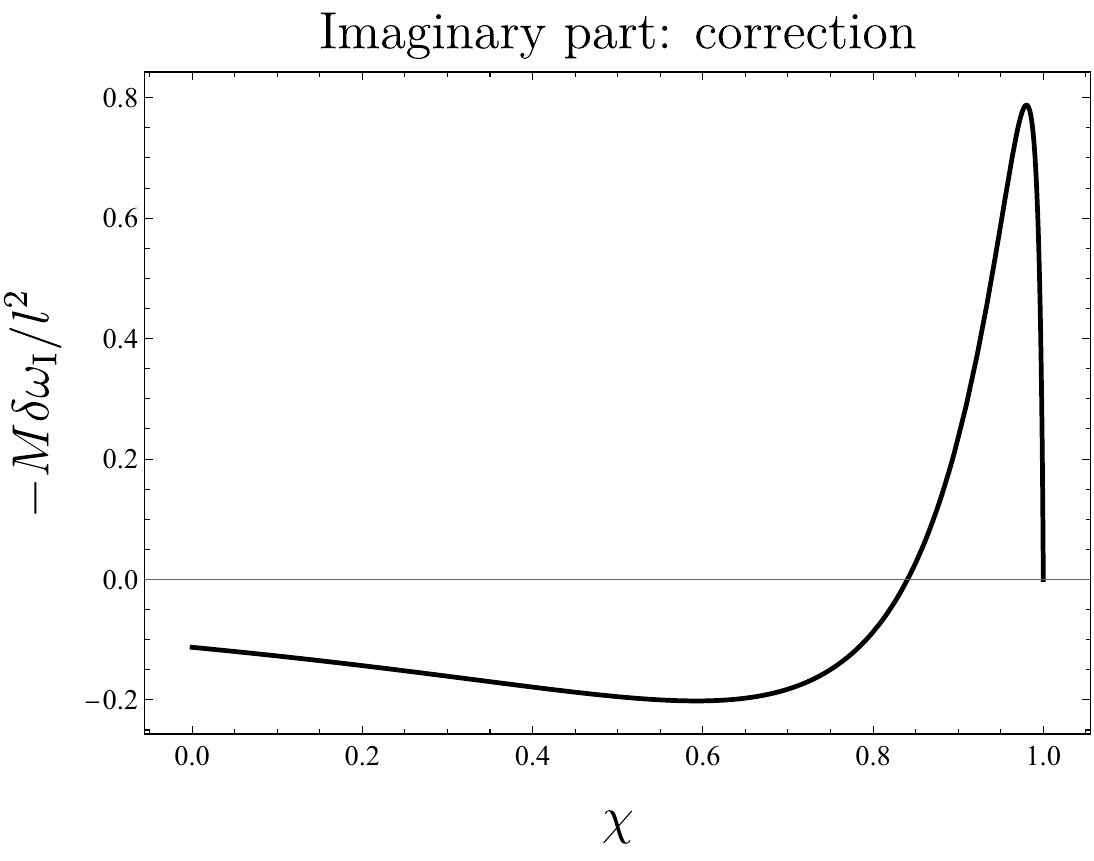}
	\caption{Kerr QNM frequencies with $\ell=m$ (top row)  and their corrections (bottom row) as a function of the spin parameter $\chi$ ranging from zero rotation up to extremality. In the case of the imaginary part we show the fundamental mode $n=0$.}\label{figure:equatorialplots}
\end{figure}

We present the QNM frequencies and their corrections in Figure~\ref{figure:equatorialplots} as a function of the rotation parameter across the full range $\chi\in [0,1]$. Note that since we are in a modified theory of gravity, it is expected that the value of $\chi$ at extremality will shift -- either larger or smaller than $\chi=1$ depending on the correction and the sign of the coupling constant. We neglect this correction as it is subleading in the eikonal limit.

One of the key features that appears as we approach $\chi=1$ is that the imaginary part of the Kerr frequency tends to zero. The real part on the other hand tends to the value $\omega_{R}=m/(2M)$. The corrections to both the real and the imaginary part vanish in the extremal limit, so this behavior of the Kerr QNMs is preserved. However, there is a sharp peak in the corrections to both the real and imaginary parts, occurring quite close to extremality. 
In fact, comparing with moderate values of the angular momentum, like $\chi\sim 0.7$, we see that the corrections grow by nearly an order of magnitude before they quickly drop to zero at extremality. The corrections to the imaginary part further exhibit a sign change. 

In the case of the imaginary part, our findings suggest that very close to extremality -- but not exactly at extremality -- there may be a competition between the Kerr result and the correction, which could potentially overcome the former. This situation would happen at the value (or range) of $\chi$ that maximizes the value of the correction while minimizing the value of the Kerr frequency. We comment on this possible situation in greater detail in section \ref{sec:arbitraryrotation}. 

\section{QNMs from the effective scalar equation}\label{sec:effective}
In order to obtain QNMs with $\ell\neq m$ and deal with the issue of non-separablity, it is more straightforward to solve directly the effective wave equation \req{effectivePhieq}. In order to write this equation in a simpler way, it is convenient to introduce the Kinnersly null tetrad for the Kerr metric, $\left\{\ell_{\mu}, n_{\mu}, m_{\mu}, \bar{m}_{\mu}\right\}$, given by 
\begin{align*}
	\ell_{\mu} dx^{\mu} &=  -dt+\frac{\Sigma}{\Delta}dr+a (1 - x^2)d\varphi\, , \\
	n_{\mu} dx^{\mu}&= \frac{\Delta}{2\Sigma} \left(-dt-\frac{\Sigma}{\Delta}dr+a (1 - x^2)d\varphi\right)\, , \\
	m_{\mu}dx^{\mu} &= \frac{\sqrt{1 - x^2}}{\sqrt{2}\, \bar{\zeta}} \left( -i a dt -\frac{\Sigma}{1 - x^2}dx+i (a^2 + r^2) d\varphi\right)\, , \\
	\bar{m}_{\mu}dx^{\mu} &=\frac{\sqrt{1 - x^2}}{\sqrt{2}\, \zeta} \left( i a dt -\frac{\Sigma}{1 - x^2}dx-i (a^2 + r^2) d\varphi\right)\, ,
\end{align*}
where 
\begin{equation}
 \zeta=r-i a x\, ,
\end{equation}
and $\Delta$ and $\Sigma$ are given by \req{eq:DeltaWsol}. Each of these vectors is null and the only non-zero contractions are 
\begin{equation}
\ell_{\mu} n^{\mu}=-1 \, \quad m_{\mu} {\bar{m}}^{\mu}=1\, .
\end{equation}
In terms of this tetrad, the Kerr metric reads 
\begin{equation}\label{gNP}
g_{\mu\nu}= -2\ell_{(\mu} n_{\nu)}+ 2m_{(\mu} \bar{m}_{\nu)}\, .
\end{equation}
Furthermore, in the Kinnersley tetrad, the Weyl tensor takes the canonical form for a Petrov D spacetime, and therefore it is given by
\begin{equation}
	\begin{split}
		W_{\mu\nu\rho\sigma}&=4\Psi_2 \left( n_{[\mu}\bar{m}_{\nu]}m_{[\rho}\ell_{\sigma]}+m_{[\mu}\ell_{\nu]}n_{[\rho}\bar{m}_{\sigma]}\right)+ 4\bar{\Psi}_2 \left( n_{[\mu}m_{\nu]}\bar{m}_{[\rho}\ell_{\sigma]}+\bar{m}_{[\mu}\ell_{\nu]}n_{[\rho}m_{\sigma]}\right)\\
		&+4(\Psi_2+\bar{\Psi}_2) \left( \ell_{[\mu}n_{\nu]}\ell_{[\rho}n_{\sigma]}+m_{[\mu}\bar{m}_{\nu]}m_{[\rho}\bar{m}_{\sigma]}\right)\\
		&-4(\Psi_2-\bar{\Psi}_2) \left( \ell_{[\mu}n_{\nu]}m_{[\rho}\bar{m}_{\sigma]}+m_{[\mu}\bar{m}_{\nu]}\ell_{[\rho}n_{\sigma]}\right)\, ,
	\end{split}
\end{equation}
where 
\begin{equation}
\Psi_2=-\frac{M}{(r-i a x)^3}\, .
\end{equation}
Since the Kerr background is Ricci flat, the Riemann tensor equals the Weyl tensor, and after some algebra, the quadratic Riemann contraction in equation \req{effectivePhieq} reads
\begin{equation}
\begin{aligned}
\tensor{R}{^{(\lambda}_{\alpha}^{\eta}_{\beta}}\tensor{R}{^{\rho|\alpha|\sigma)\beta}}=&- 18(\Psi_2-\bar{\Psi}_2)^2  m^{(\lambda}\bar{m}^{\eta} n^{\rho} \ell^{\sigma)}\\
&+ 6(\Psi_2+\bar{\Psi}_2)^2 \left[ m^{(\lambda}\bar{m}^{\eta} m^{\rho} \bar{m}^{\sigma)}+n^{(\lambda}\ell^{\eta} n^{\rho} \ell^{\sigma)}+ m^{(\lambda}\bar{m}^{\eta} n^{\rho} \ell^{\sigma)}\right]\, .
\end{aligned}
\end{equation}
Taking then into account that the Laplacian is given by
\begin{equation}
\nabla^2=-2\ell^{(\mu}n^{\nu)}\nabla_{\mu}\nabla_{\nu}+2m^{(\mu}\bar{m}^{\nu)}\nabla_{\mu}\nabla_{\nu}\, ,
\end{equation}
we can write \req{effectivePhieq} as
\begin{equation}
\begin{aligned}
\nabla^2\Phi+64\alpha \Big[&72 \Psi_{2}\bar{\Psi}_{2}m^{(\alpha}\bar{m}^{\beta} m^{\mu}\bar{m}^{\nu)}\nabla_{\alpha}\nabla_{\beta}\nabla_{\mu}\nabla_{\nu}\Phi-36 \Psi_{2}\bar{\Psi}_{2}m^{(\alpha}\bar{m}^{\beta)} \nabla_{\alpha}\nabla_{\beta}\nabla^2\Phi\\
&+\frac{3}{2}(\Psi_2+\bar{\Psi}_2)^2\left(\nabla^2\right)^2\Phi\Big]=0\, ,
\end{aligned}
\end{equation}
plus terms with a lower number of derivatives of $\Phi$ arising from the commutation of covariant derivatives that we neglect as they are subleading in the eikonal limit.  Now, all the higher-derivative terms containing a Laplacian are subleading, since $\nabla^2\Phi=\mathcal{O}(\alpha)$. Therefore, at first order in $\alpha$, the equation reduces to 
\begin{equation}\label{Phieqsimp}
\nabla^2\Phi+4608\alpha  \Psi_{2}\bar{\Psi}_{2}m^{(\alpha}\bar{m}^{\beta} m^{\mu}\bar{m}^{\nu)}\nabla_{\alpha}\nabla_{\beta}\nabla_{\mu}\nabla_{\nu}\Phi=0\, .
\end{equation}
One can check that this equation is non-separable and so in order to solve it, we expand $\Phi$ using the spheroidal harmonics as a basis of angular functions, a technique already employed in previous literature \cite{Cano:2020cao,Ghosh:2023etd}. We thus write
\begin{equation}
\Phi=e^{-i\omega t+im\varphi}\sum_{\ell=m}^{\infty}S_{\ell m}(x; a\omega) R_{\ell m}(r)\, ,
\end{equation}
where the spheroidal harmonics satisfy the equation
\begin{equation}\label{spheroidaleq}
\frac{d}{dx}\left[(1-x^2)\frac{dS_{\ell m}}{dx}\right]+\left(A_{\ell m}(a\omega)+a^2 \omega^2 x^2-\frac{m^2}{1-x^2}\right)S_{\ell m}=0\, ,
\end{equation}
where $A_{\ell m}(a\omega)$ are the angular separation constants. The Laplacian then yields
 \begin{equation}
\nabla^2\Phi=e^{-i\omega t+im\varphi}\frac{1}{\Sigma}\sum_{\ell=m}^{\infty}S_{\ell m} \mathcal{D}^2_{\ell m}R_{\ell m}\, ,
\end{equation}
where 
\begin{equation}
 \mathcal{D}^2_{\ell m}R_{\ell m}=\frac{d}{dr}\left(\Delta\frac{dR_{\ell m}}{dr}\right)+\frac{V}{\Delta}R_{\ell m}\, ,
\end{equation}
and $V$ is the scalar Teukolsky potential
\begin{equation}\label{V0}
V=\left[\omega\left(r^2+a^2\right) -am\right]^2-\lambda_{\ell m}\Delta\, ,
\end{equation}
where for convenience we have defined
\begin{equation}
\lambda_{\ell m}\equiv A_{\ell m}-2m a\omega+(a\omega)^2\, .
\end{equation}
On the other hand, the fourth-derivative operator gives us the following result in the eikonal limit
\begin{equation}\label{fourth}
m^{(\alpha}\bar{m}^{\beta} m^{\mu}\bar{m}^{\nu)}\nabla_{\alpha}\nabla_{\beta}\nabla_{\mu}\nabla_{\nu}\Phi=e^{-i\omega t+im\varphi}\frac{1}{4\Sigma^2}\sum_{\ell=m}^{\infty}\lambda_{\ell m}^2S_{\ell m} R_{\ell m}+\ldots\, ,
\end{equation}
where we have used the equation \req{spheroidaleq} to simplify the result. The ellipsis denote terms with a lower number of derivatives acting on $\Phi$, which are subleading in the eikonal limit. 
The full equation \req{Phieqsimp} then becomes
\begin{equation}
\sum_{\ell=m}^{\infty}S_{\ell m}\left[\mathcal{D}^2_{\ell m}R_{\ell m}+1152\alpha M^2\frac{\lambda_{\ell m}^2}{\Sigma^4}R_{\ell m}\right]=0\, .
\end{equation}
We now project it onto the spheroidal harmonics using the orthogonality property
\begin{equation}
\int_{-1}^{1} dx S_{\ell m}(x; a\omega) S_{\ell' m}(x; a\omega)=2\pi \delta_{\ell \ell'}\, ,
\end{equation}
and we get a system of radial equations

\begin{equation}\label{Radialsystem}
\mathcal{D}^2_{\ell m}R_{\ell m}+1152\alpha M^2\sum_{\ell'=m}^{\infty}R_{\ell' m}\lambda_{\ell' m}^2\int_{-1}^{1} dx\frac{S_{\ell m}(x; a\omega) S_{\ell' m}(x; a\omega)}{2\pi \Sigma^4}=0\, .
\end{equation}
Finally, we observe that the solution will have a leading mode, say $\ell=\ell_0$, such that $R_{\ell_0 m}=\mathcal{O}(1)$ and $R_{\ell\neq\ell_0 m}=\mathcal{O}(\alpha)$. This is because in the GR case, the solution consists of a single $\ell$-mode, but the corrections will in general turn on all the other harmonics. Since all the radial functions with $\ell\neq \ell_{0}$ are of order $\alpha$, the sum in \req{Radialsystem} only contains a single term at first order in $\alpha$: the one corresponding to $\ell=\ell_0$. The radial equation for $\ell=\ell_0$ is therefore decoupled, and we can write it as
\begin{equation}\label{mastermodTeuk}
\Delta\frac{d}{dr}\left(\Delta\frac{dR_{\ell m}}{dr}\right)+\left(V+\hat{\alpha}  \delta V\right)R_{\ell m}=0\, ,
\end{equation}
where $\hat\alpha$ is the dimensionless coupling constant in \req{hatalpha}, 
and the correction to the potential reads
\begin{equation}\label{deltaV0}
\delta V=1152 M^8\Delta \lambda_{\ell m}^2\int_{-1}^{1} dx\frac{S_{\ell m}(x; a\omega)^2}{2\pi (r^2+a^2x^2)^4}\, .
\end{equation}

\subsubsection*{Eikonal limit}
Before continuing, we remark on a few details about the eikonal limit. The eikonal regime is defined by the limit $\ell\to\infty$, $m\to\infty$ with $m/\ell$ fixed. At leading order in $\ell$ (and in GR), this leads to the scalings $\omega\sim \ell$, $A_{\ell m}\sim \ell^2$. In fact, as shown by \cite{Yang:2012he}, the angular separation constants are very approximately given by
\begin{equation}\label{Almeik}
A_{\ell m}(a\omega)\approx L^2-\frac{a^2\omega^2}{2}\left(1-\frac{m^2}{L^2}\right)\, ,
\end{equation}
where 
\begin{equation}
L=\ell+\frac{1}{2}\, ,
\end{equation}
which we will use as the expansion parameter from now on instead of $\ell$. We remark that \req{Almeik} is not exact in $a\omega$, but its error is smaller than $0.2\%$ for all the values of $a\omega/L$ that occur in Kerr QNMs \cite{Yang:2012he}. It is also convenient to introduce the quantity
\begin{equation}
\mu=\frac{m}{L}\in (-1,1)\, ,%\quad \Omega=\frac{\omega}{L}\, ,\quad  \sigma=\frac{A_{\ell m}}{L^2}\approx 1-\frac{a^2\Omega^2}{2}(1-\mu^2)\, .
\end{equation}
which we will use throughout. From these scalings, it follows that the Kerr potential grows quadratically, $V\sim L^2$, but its correction grows faster $\delta V\sim L^4$. This means that the corrections eventually overcome the GR prediction for sufficiently large $L$,  representing a breakdown of the EFT. Thus, we must consider a regime in which $L$ is large, so that the eikonal description is accurate, but not \emph{too large}, so that we remain within the regime of validity of the EFT.  This ``Goldilocks'' regime corresponds to
\begin{equation}\label{eq:validity}
1\ll L \ll |\hat{\alpha}|^{-1/6}\, ,
\end{equation}
where the second condition ensures that the wavelength of the perturbation is larger than the length scale of new physics $|\alpha|^{1/6}$. We observe that, in order for the background spacetime to be valid within the EFT, we must require $|\hat{\alpha}|\ll 1$, and thus \req{eq:validity} can always be satisfied. 

On another ground, we have verified that the subleading terms that we dropped in \req{fourth}, do not introduce any $\mathcal{O}(L^3)$ terms in the radial equation \req{mastermodTeuk}. Therefore, the correction to the potential  \req{deltaV0} is actually valid up to order $\mathcal{O}(L^2)$. 

\subsection{Correction to the Teukolsky potential}
We have reduced the problem of analyzing eikonal gravitational perturbations to the single radial equation \req{mastermodTeuk}. However, this still entails a final difficulty, since the correction to the potential involves an integral of spheroidal harmonics, 
\begin{equation}\label{Ilm}
I_{\ell m}=\frac{1}{2\pi}\int_{-1}^{1} dx\frac{S_{\ell m}(x; a\omega)^2}{(r^2+a^2x^2)^4}\, ,
\end{equation} 
that become highly oscillatory in the eikonal limit, making it intractable numerically. Therefore, we need to simplify this result and reduce it to a more manageable expression. 
Here we follow a similar strategy to the one introduced in \cite{Cano:2024bhh} in order to deal with this kind of integrals. 
The idea is to modify the integrand  of \req{Ilm} with a ``gauge term'' that does not affect the result of the integral.  To achieve this, we use the following result.

Let $h: [-1,1]\to \mathbb{R}$ be a differentiable function satisfying $h(1)=h(-1)=0$ and $\mathcal{F}[h]$ the third-order differential operator defined by 
\begin{eqnarray}\label{F_definition}\nonumber
\mathcal{F}[h]&=&-\frac{1}{2}(1-x^2)h^{(3)}(x)+\frac{2}{(1-x^2)}\left( m^2-a^2 \omega^2 x^2 (1-x^2)-1-A_{\ell m}(1-x^2) \right)h'(x)\\
 &&-\frac{2}{(1-x^2)^2}\left[ A_{\ell m}x(1-x^2)- x \left(2 m^2 -a^2 \omega^2  (1-x^2)\right)+2x\right]h(x)\, .
\end{eqnarray}
Then, the following integral vanishes identically, 
\begin{equation}\label{gauge_condition}
\int_{-1}^{1} dx (S_{\ell m})^2 \mathcal{F}[h]= 0\, .
\end{equation}
%%%%%%%%%%%%%%%%%%%%%%%%%%%%%%%%%%%%
%%%%%%%%%%%%%%%%%%%%%%%%%%%%%%%%%%%%
In order to prove this, we integrate by parts the term with three derivatives, so that we get an integrand that depends only on $h'(x)$ and $h(x)$. Then, we arrive at
\begin{equation}\label{gauge_inv_appendix}
\begin{split}
&\int_{-1}^{1} dx (S_{\ell m})^2 \mathcal{F}[h]=g(h(x))\biggr\rvert_{-1}^{1}\\
& -\int_{-1}^{1}{\rm{d}}x\left(h' (x)S_{\ell m}-2 h(x) S'_{\ell m}+\frac{2 x}{(1-x^2)}h(x) S_{\ell m} \right)\left( \mathcal{D}_x^2 +A_{\ell m}+a^2\omega^2x^2-\frac{m^2}{1-x^2}\right)S_{\ell m}\, ,
\end{split}
\end{equation}
where we have the boundary term
\begin{eqnarray}\label{gauge_inv_appendix_2}\nonumber
g(h(x))&=& -h(x) S_{\ell m}\left( \mathcal{D}_x^2 +A_{\ell m}+a^2\omega^2x^2-\frac{m^2}{1-x^2}\right)S_{\ell m}\\\nonumber
&& -(1-x^2)\left[-S_{\ell m} (h(x) S_{\ell m}')'+h(x){S_{\ell m}'}^2+\frac{1}{2}S_{\ell m}^2h''(x)\right]\\
&&-S_{\ell m}^2\left(x h(x)'+\frac{(1+x^2)}{1-x^2}h(x)\right)\, ,
\end{eqnarray}
and where we defined the differential operator $\mathcal{D}_x^2 S\equiv \frac{d}{dx}\left[ (1-x^2)\frac{d S}{dx}\right]$.  The integrand  in the second line of \req{gauge_inv_appendix} vanishes identically because it is proportional to the equation of the spheroidal harmonics \req{spheroidaleq}. For the same reason, the first line of the boundary piece \req{gauge_inv_appendix_2} also vanishes. Furthermore, the second line of \req{gauge_inv_appendix_2} is proportional to $(1-x^2)$, meaning that when evaluated at the end points it is equal to zero, given that that the functions $h(x)$ and $S_{\ell m}(x)$ are regular at $x=\pm 1$. Lastly, it is possible to show that the third line in \req{gauge_inv_appendix_2} vanishes as well for $x\to \pm 1$, since $\lim_{x\to \pm 1} (1+x^2)h(x)/(1-x^2)=\mp h'(\pm 1)$. This follows from the fact that that $h(x)$ is a differentiable function that satisfies $h(\pm 1)=0$. Therefore, we conclude the proof of \req{gauge_condition}.
%%%%%%%%%%%%%%%%%%%%%%%%%%%%%%%%%%%%
%%%%%%%%%%%%%%%%%%%%%%%%%%%%%%%%%%%%

As a consequence of \req{gauge_condition}, we can equivalently consider the integral 
\begin{equation}\label{Ilm2}
I_{\ell m}=\frac{1}{2\pi}\int_{-1}^{1} dxS_{\ell m}(x; a\omega)^2\left[(r^2+a^2x^2)^{-4}+\mathcal{F}[h]\right]\, ,
\end{equation} 
and our idea is to choose a function $h(x)$ such that the integrand simplifies. In fact, if we can find a function $h$ such that
\begin{equation}\label{diffh}
(r^2+a^2x^2)^{-4}+\mathcal{F}[h]=\kappa\, ,
\end{equation}
for a constant $\kappa$, then it would follow that $I_{\ell m}=\kappa$, due to the normalization of the spheroidal harmonics. Of course, $\kappa$ cannot be arbitrary, since otherwise the result of the integral would be arbitrary. This constant is fixed by demanding that $h(x)$ --- which now is a solution of the differential equation \req{diffh} --- is regular in the interval $x\in [-1,1]$ and satisfies $h(\pm 1)=0$.  
While obtaining $\kappa$ analytically is not possible in general (because \req{diffh} is a third-order differential equation), in the eikonal limit this constant can in fact be calculated analytically. 

To order $\mathcal{O}(\ell^2)$ in the eikonal limit $\ell\to\infty$, the operator $\mathcal{F}[h]$ becomes
\begin{equation}
\begin{aligned}
\mathcal{F}[h]&=\frac{2}{(1-x^2)}\left( m^2-a^2 \omega^2 x^2 (1-x^2)-A_{\ell m}(1-x^2) \right)h'(x)\\
 &+\frac{2x}{(1-x^2)^2}\left[2 m^2 -a^2 \omega^2 (1-x^2) -A_{\ell m}(1-x^2)\right]h(x)\, .
\end{aligned}
\end{equation}
Thus, the equation \req{diffh} becomes of first order, and now an explicit solution exists, which reads
\begin{equation}\label{hsol1}
h(x)=\frac{(1-x^2)}{\sqrt{Z(x)}}\int_{x_1}^{x} dx'\frac{\kappa-(r^2+a^2x'^2)^{-4}}{\sqrt{Z(x')}}\, ,
\end{equation}
where
\begin{equation}
Z(x)=A_{\ell m}(1-x^2)-m^2+(a\omega)^2x^2(1-x^2)\, ,
\end{equation}
and the limit of integration $x_{1}$ is an integration constant.  Now we have to take into account that, in order for $h$ to be an allowable transformation of the integral \req{Ilm2}, it must be smooth in the interval $x\in[-1,1]$, and vanish at $x=\pm 1$.  From \req{hsol1} we see  that the latter condition is generically satisfied, but the former is not: $h(x)$ is in general singular at the points in which $Z(x)=0$.  
In order to determine the roots of $Z(x)$, we use the approximate expression for the angular separation constants \req{Almeik}, and in addition we take $\omega$ to be the real part of the Kerr QNM frequency, $\omega_{R}$, since the imaginary part is subleading in the eikonal limit.  Then we have
\begin{align}
Z(0)&=A_{\ell m}-m^2\approx (L^2-m^2)\left(1-\frac{(a\omega_{R})^2}{2L^2}\right)> 0\,, \\
Z(1)&=-m^2\le 0\, ,
\end{align}
where in the first equation we used \req{Almeik} and took into account that all Kerr QNM frequencies satisfy $a\omega_{R}/L \le 1/2$ \cite{Yang:2012he}. Since $Z(x)\to -\infty$ for $x\to\pm \infty$, these inequalities imply that $Z(x)$ always has two symmetric roots at $x=\pm x_0$, with $x_0\in [0,1]$ given by
\begin{equation}\label{x0}
x_0=\frac{1}{\sqrt{2}a\omega}\left[-A_{\ell m}+(a\omega)^2+\sqrt{(A_{\ell m}+(a\omega)^2)^2-4m^2 a^2\omega^2}\right]^{1/2}\, .
\end{equation}
Then, in order for \req{hsol1} to be regular at $x=\pm x_0$, it is necessary for the integral that appears in this expression to vanish for both $x=\pm x_0$.  Straightforwardly, this fixes $x_{1}=-x_0$ and leads to the condition
\begin{equation}\label{hsol2}
\int_{-x_0}^{x_0} dx\frac{\kappa-(r^2+a^2x^2)^{-4}}{\sqrt{Z(x)}}=0\, ,
\end{equation}
which determines the value of $\kappa$, and consequently, of the integral \req{Ilm}:
\begin{equation}\label{kappa1}
I_{\ell m}=\kappa=\frac{\int_{-x_0}^{x_0} dx(r^2+a^2x^2)^{-4}Z(x)^{-1/2}}{\int_{-x_0}^{x_0} dx Z(x)^{-1/2}}\, .
\end{equation}
Finally, the integral can be simplified by performing the change of variable $x=x_0 \sin u$ and we get
\begin{equation}\label{Ilm3}
I_{\ell m}=\frac{1}{K(-q)}\int_{0}^{\pi/2} \frac{du}{(r^2+a^2x_0^2\sin^2 u)^{4}\sqrt{1+q \sin^2 u}}\, ,
\end{equation}
where 
\begin{equation}\label{qsol}
q=\frac{x_0^2 (a\omega)^2}{A_{\ell m}-(1-x_0^2)(a\omega)^2}\, ,
\end{equation}
and $K(-q)$ is the elliptic integral of the first kind, 
\begin{equation}
K(-q)=\int_{0}^{\pi/2} \frac{du}{\sqrt{1+q \sin^2 u}}\, . 
\end{equation}
Therefore, our final expression for the potential reads
\begin{equation}\label{deltaV1}
\delta V=\frac{1152 M^8\Delta \lambda_{\ell m}^2}{r^8 K(-q)}\int_{0}^{\pi/2} \frac{du}{\left(1+\frac{a^2x_0^2}{r^2}\sin^2 u\right)^{4}\sqrt{1+q \sin^2 u}}\, .
\end{equation}

\subsection{Solution via the WKB approximation}
Performing the change of variable $R_{\ell m}=\frac{\psi}{\sqrt{r^2+a^2}}$, we can rewrite our master equation \req{mastermodTeuk} as
\begin{equation}\label{WKBform}
\frac{d^2\psi}{dr_{*}^2}+U\psi=0\, ,
\end{equation}
where
\begin{equation}
\frac{d}{dr_{*}}=\frac{\Delta}{r^2+a^2}\frac{d}{dr}\, ,\quad  U=\frac{V+\hat{\alpha}\delta V}{(r^2+a^2)^2}\, ,
\end{equation}
and where we are neglecting terms that are subleading in the eikonal limit. The equation \req{WKBform} can then be solved with the WKB method, which becomes exact in the eikonal limit.  The leading-order WKB approximation yields a ``Bohr-Sommerfeld'' quantization condition that reads \cite{Iyer:1986np} 
\begin{equation}
\frac{U}{\sqrt{2\partial_{r_{*}}^2U}}\bigg|_{r_0}=-i\left(n+\frac{1}{2}\right)\, ,
\end{equation}
where $r_0$ represents the minimum of the potential, $\partial_{r} U(r_0)=0$. 
Taking into account that the real part of the frequency is dominant in the eikonal limit, $\omega_{R}\sim L$, while the imaginary part is subleading, $\omega_{I}\sim L^{0}$, this leads to the following prescription. The real part is determined by extremization conditions
\begin{equation}\label{WKBR}
U\Big|_{r_0,\omega_{R}}=\frac{dU}{dr}\bigg|_{r_0,\omega_{R}}=0\, ,
\end{equation}
while the imaginary part is given by 
\begin{equation}\label{WKBI}
\omega_{I}=-\left(n+\frac{1}{2}\right)\frac{\sqrt{2\partial_{r_{*}}^2U}}{\partial_{\omega}U}\bigg|_{r_0,\omega_{R}}=-\left(n+\frac{1}{2}\right)\frac{\Delta\sqrt{2\partial_{r}^2(V+\hat\alpha\delta V)}}{\partial_{\omega}(V+\hat\alpha \delta V)}\bigg|_{r_0, \omega_{R}}\, .
\end{equation}
We remark that in the computation of $\partial_{\omega} U$ we have to take into account the dependence of the angular separation constants on the frequency --- we will use the approximate formula \req{Almeik}. Let us also point out that \req{WKBI} is the same as our final formula for the damping factor $\gamma$ \req{gammamaster} arising from the geometric-optics approach, once we take into account that $\mathcal{U}=-U|_{m=\ell}$. 

\begin{figure}[t!]
  	\centering
  	\includegraphics[width=0.99\textwidth]{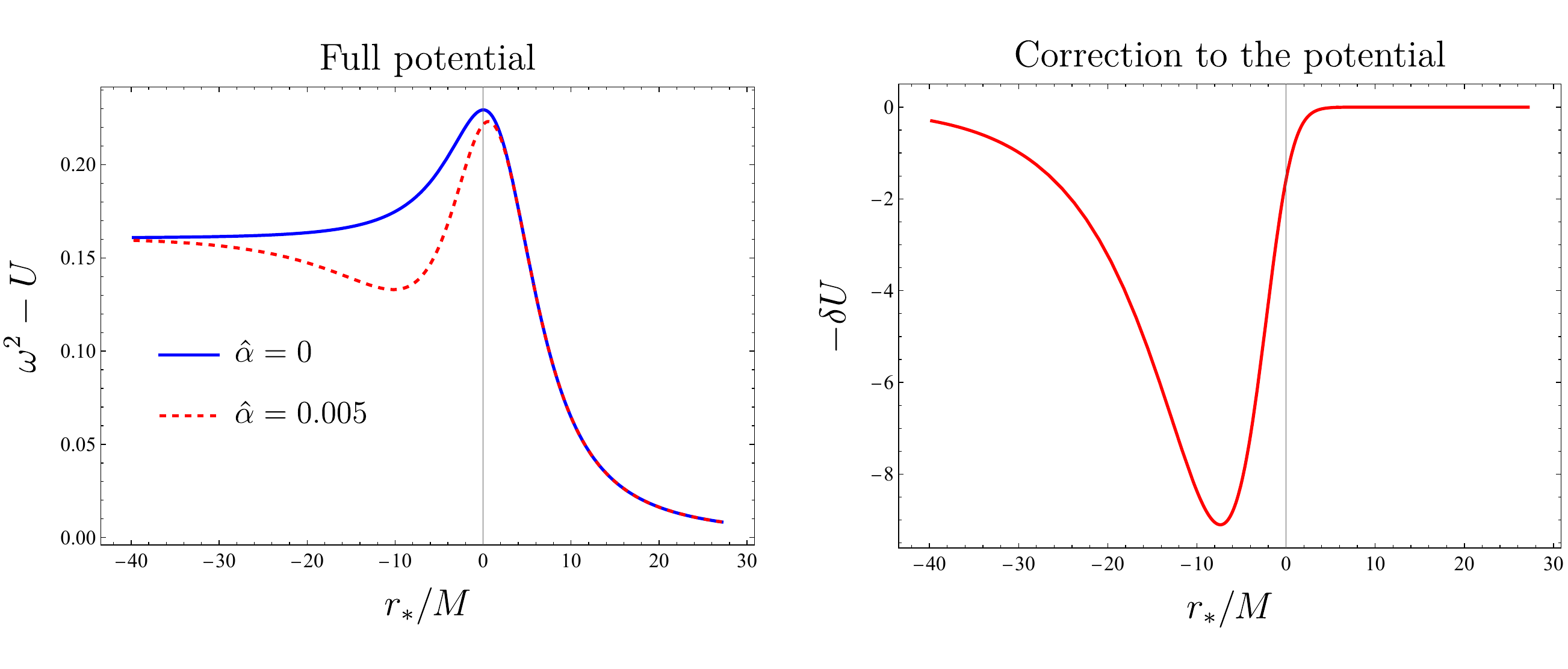}
  	\caption{Potential as a function of the tortoise coordinate. \textbf{Left}:  full potential for $\hat\alpha=0$ (GR) and for $\hat\alpha=0.005$. We plot  $\omega^2-U$, which vanishes at infinity and represents the intuitive definition of a potential. The corrections introduce in this case a local minimum, but it disappears for smaller values of $\hat\alpha$. \textbf{Right}: correction to the potential, which exhibits a minimum quite close to the horizon.  The plots correspond to $\chi=0.99$, $\mu=1/4$, and  $\omega=\omega_{R}^{\rm Kerr}$. }
  	\label{fig:potential}
  \end{figure}

Another important remark about the WKB approximation is that it only works on single-peak potentials. We observe that the correction $\delta V$ has a peak\footnote{It can be either a peak or a valley depending on the sign of the coupling constant $\alpha$.} in a different location than the peak of $V$. This means that if $\hat{\alpha}$ is large enough, the full potential may develop a second peak or a local minimum, which would greatly affect the QNM spectrum. Typically these situations require large corrections, so they happen outside the regime of validity of the EFT.
However, we have observed that even values of $\hat\alpha$ much smaller than one can produce a second peak or a valley (see Figure~\ref{fig:potential}). It would be interesting to understand if these effects can appear consistently within the EFT regime, but we leave this question for future work.      Here we will restrict  to the regime in which $\hat\alpha$ is small enough so that the shape of the potential is qualitatively unchanged and we can apply the WKB approximation.

 For small enough $\hat\alpha$ we can consider an expansion of the frequency and of $r_0$ around their Kerr values, 
\begin{align}
\omega_{R}=\omega_{R}^{\rm Kerr}+\hat\alpha \delta\omega_{R}\, ,\quad \omega_{I}=\omega_{I}^{\rm Kerr}+\hat\alpha \delta\omega_{I}\, ,\quad r_{0}=r_{0}^{\rm  Kerr}+\hat\alpha \delta r_{0}\, .
\end{align}
Linearizing \req{WKBR} and \req{WKBI} at first order in $\hat\alpha$, it is straighforward to obtain explicit expressions for the shifts, 
\begin{align}\label{r0gen}
\delta r_{0}&=\frac{\delta V \partial_{r}\partial_{\omega} V-\partial_{r}\delta V\partial_{\omega}V}{\partial_{\omega} V\partial_{r}^2 V}\bigg|_{r_{0}^{\rm Kerr}, \omega_{R}^{\rm Kerr}}\, ,\\\label{deltawRgen}
\delta\omega_{R}&=-\frac{\delta V}{\partial_{\omega} V}\bigg|_{r_{0}^{\rm Kerr}, \omega_{R}^{\rm Kerr}}\, ,\\\label{deltawIgen}
\frac{\delta\omega_{I}}{\omega^{\rm Kerr}_{I}}&=\frac{\partial_{r}^2\delta V}{2\partial_{r}^2 V}-\frac{\partial_{\omega}\delta V}{\partial_{\omega} V}+\delta \omega_{R}\frac{\partial}{\partial\omega}\log\left(\frac{\sqrt{\partial_{r}^2V}}{\partial_{\omega}V}\right)+\delta r_{0}\frac{\partial}{\partial r}\log\left(\Delta\frac{\sqrt{\partial_{r}^2V}}{\partial_{\omega}V}\right)\bigg|_{r_{0}^{\rm Kerr}, \omega_{R}^{\rm Kerr}}\, .
\end{align}
In order to evaluate these expressions, we only need to plug in the values of $r_{0}^{\rm Kerr}$, $\omega_{R}^{\rm Kerr}$ and $\omega_{I}^{\rm Kerr}$. These were found explicitly in \cite{Yang:2012he} and we reproduce those results below.  In order to shorten the formulas, we introduce the dimensionless quantities  
\begin{equation}
\chi=\frac{a}{M}\, ,\quad z=\frac{r_{0}^{\rm  Kerr}}{M}\, .
\end{equation}
The values of $\omega_{R}^{\rm Kerr}$ and $r_{0}^{\rm  Kerr}=z M$ follow from solving $V=\partial_{r} V=0$ for the Kerr potential \req{V0}. From the equation $\partial_{r}(V/\Delta)=0$, we find that $\omega_{R}^{\rm Kerr}$ is given by 
\begin{equation}\label{wRKerr}
M\omega_{R}^{\rm Kerr}=L \frac{(1-z)\mu \chi}{(z-3)z^2+(z+1)\chi^2}\, .
\end{equation}
Inserting this back into $V=0$ yields a polynomial equation that fixes $z$ as a function of $\chi$,
\begin{equation}\label{zchirel}
\begin{aligned}
&2z^4(z-3)^2+4z^2\left[(1-\mu^2)z^2-2z-3(1-\mu^2)\right]\chi^2\\
&+(1-\mu^2)\left[(2-\mu^2)z^2+2(2+\mu^2)z+2-\mu^2\right]\chi^4=0\, .
\end{aligned}
\end{equation}
Evaluating \req{WKBI} for $\hat\alpha=0$, and using  \req{wRKerr} and \req{zchirel}, we find that the imaginary part can be expressed exactly as found in \cite{Yang:2012he}, 
\begin{equation}\label{wIKerr}
M\omega_{I}^{\rm Kerr}=-\frac{\left(n+\frac{1}{2}\right)(z^2-2z+\chi^2)\sqrt{4(6z^2\Omega^2-1)+2\chi^2\Omega^2(3-\mu^2)}}{2z^4\Omega-4z\chi\mu+\chi^2 z \Omega\left[z(3-\mu^2)+2(1+\mu^2)\right]+\chi^4\Omega(1-\mu^2)}\Big|_{\Omega=\frac{M\omega_{R}^{\rm Kerr}}{L}}\, ,
\end{equation}
where we have assumed the approximate result for the angular separation constants \req{Almeik}. 

\section{Corrections to the QNM frequencies}\label{sec:QNMs}
In this section we analyze in detail the shifts in the QNM frequencies, given by \req{deltawRgen} and \req{deltawIgen}. These are functions of two variables: the dimensionless angular momentum of the black hole $\chi$, and the ratio $\mu=m/L$. We also note that the real and imaginary parts scale as $\delta\omega_{R}\sim L^3$ and $\delta\omega_{I}\sim L^2$. 
\subsection{Analytic expressions}
Although our results \req{deltawRgen} and \req{deltawIgen} are in principle analytic, these formulas are rather involved since the potential is given by the integral \req{deltaV1} and the dependence on $\chi$ and $\mu$ is rather implicit. 
For illustrative purposes, we consider here two instances where simple analytic expressions can be obtained: small rotation $\chi\ll 1$ and $\mu$ close to 1. In the latter case, we observe that the value of $x_0$ --- defined in \req{x0} ---  vanishes when $\mu\to 1$. This means that in both cases we have $q\ll 1$ and $a x_0/r\ll 1$ and therefore we can expand the integral in \req{deltaV1} as 
\begin{equation}
\delta V=\frac{1152 M^8\Delta \lambda_{\ell m}^2}{r^8}\left[1-2\frac{a^2x_0^2}{r^2}\left(1-\frac{q}{8}\right)+\frac{15 a^4x_0^4}{4 r^4}+\ldots\right]\, .
\end{equation}
In addition, in both cases the equation \req{zchirel} can be solved analytically to find the explicit relationship $z(\mu,\chi)$. 

For small rotation, we find
\begin{align}
M\omega_{R}^{\rm Kerr}&=L\left[\frac{1}{3 \sqrt{3}}+\frac{2 \mu  \chi }{27}+\frac{\left(15 \mu ^2+7\right) \chi ^2}{324 \sqrt{3}}+\ldots\right]+\mathcal{O}(L^{-1})\, ,\\
M\omega_{I}^{\rm Kerr}&=\left(n+\frac{1}{2}\right)\left[-\frac{1}{3 \sqrt{3}}+\frac{2 \chi ^2}{81 \sqrt{3}}+\ldots\right]+\mathcal{O}(L^{-2})\,,
\end{align}
for the Kerr frequencies, while the corrections read
\begin{align}
M\delta\omega_{R}&=L^3\left[-\frac{64}{2187 \sqrt{3}}-\frac{256 \mu  \chi }{6561}-\frac{80 \left(51 \mu ^2+23\right) \chi ^2}{59049 \sqrt{3}}+\ldots\right]+\mathcal{O}(L)\, ,\\
M\delta\omega_{I}&=L^2\left(n+\frac{1}{2}\right)\left[-\frac{1280}{6561 \sqrt{3}}-\frac{7936 \mu  \chi }{59049}+\frac{64 \left(301 \mu ^2-718\right) \chi ^2}{177147 \sqrt{3}}+\ldots\right]+\mathcal{O}(L^{0})\, .
\end{align}
Higher-order terms in the $\chi$-expansion can easily be found, although they are not particularly illuminating. 

The expansion around $\mu=1$ is more interesting since it is valid for arbitrary rotation. The results in this case are conveniently expressed in terms of the quantity $\rho$ in \req{rhosol}, which represents the (dimensionless) radius of the equatorial photon ring for Kerr black holes. 
We find that the Kerr QNM frequencies read
\begin{align}
M\omega_{R}^{\rm Kerr}&=L\left[\frac{2}{\sqrt{\rho } (\rho +3)}-\frac{(\mu -1) (\rho -3) \left(\rho ^2+12 \rho +3\right)}{4 \left(\rho ^{3/2} (\rho +3)^2\right)}+\ldots\right]+\mathcal{O}(L^{-1})\, ,\\
M\omega_{I}^{\rm Kerr}&=\left(n+\frac{1}{2}\right)\left[-\frac{\sqrt{3} (\rho -1)}{\rho  (\rho +3)}+\frac{(\mu -1) (\rho -3)^4 (\rho -1)^2}{32 \sqrt{3} \rho ^3 (\rho +3)^2}+\ldots\right]+\mathcal{O}(L^{-2})\, ,
\end{align}
while the corrections are given by
\begin{align}\notag
M\delta\omega_{R}&=L^3\left[-\frac{4608 (\rho -1)}{\rho ^{11/2} (\rho +3)^4}\right.\\
&\left.+\frac{192 (\mu -1) \left(\rho-1\right) (\rho -3) \left(4 \rho ^4+77 \rho ^3+180 \rho ^2-405 \rho -216\right)}{\rho ^{15/2} (\rho +3)^5}+\ldots\right]+\mathcal{O}(L)\, ,\\\notag
M\delta\omega_{I}&=L^2\left(n+\frac{1}{2}\right)\bigg[-\frac{192 \sqrt{3} (\rho -1)^2 \left(35 \rho ^2+22 \rho -141\right)}{\rho ^7 (\rho +3)^4}\\\notag
&+\frac{2 \sqrt{3} (\mu -1) (\rho -3) (\rho -1)^2}{\rho ^9 (\rho +3)^5} \left(777 \rho ^6+15416 \rho ^5+28297 \rho ^4-162792 \rho ^3\right.\\
&\left.-251397 \rho ^2+368496 \rho +233523\right)+\ldots\bigg]+\mathcal{O}(L^{0})\, .
\end{align}
For $\mu=1$, these expressions exactly match our results in section \ref{sec:geometricoptics} --- see \req{omegaRgeom} and \req{omegaImgeom} --- and therefore we check that the geometric optics approach correctly identifies the QNM frequencies also in extensions of GR.  

\subsection{Comparison with the full modified Teukolsky equation}\label{subsec:comparison}
A crucial test to check the validity of our approach is to compare our results with the exact corrections to the Kerr QNM frequencies obtained through the modified Teukolsky equation. In \cite{Cano:2024ezp}, results were obtained for up to $\ell=4$ modes and moderate rotation $\chi\lesssim 0.8$ in the general EFT extension of GR. 
We can compare our eikonal prediction with those results in the regime of moderate rotation.

\begin{figure}[t!]
  	\centering
  	\includegraphics[width=0.99\textwidth]{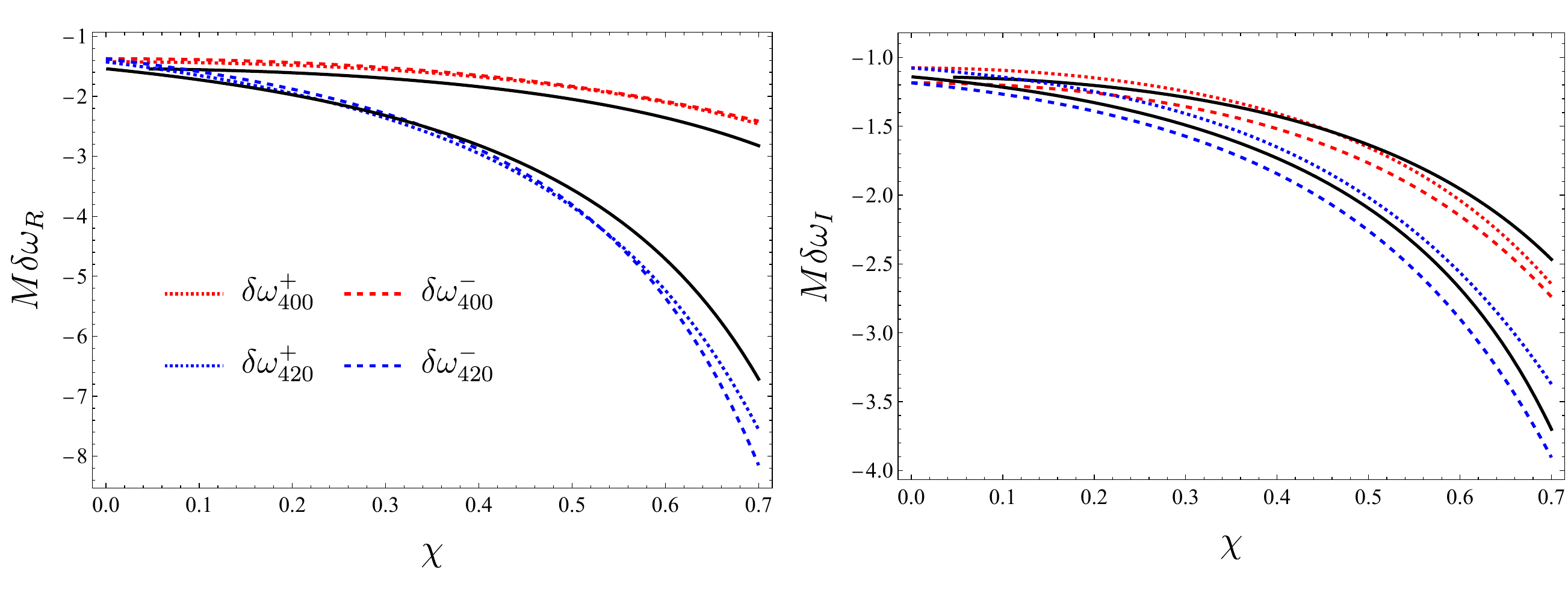}
  	\caption{Corrections to the Kerr QNM frequencies for $(\ell,m,n)=(4,0,0), (4,2,0)$. The colored dotted and dashed lines are the prediction from the modified Teukolsky equation, and they show a mild isospectrality breaking of the two families of modes $\delta\omega^{+}$ and $\delta\omega^{-}$. The solid black lines are the prediction from our eikonal computation.}
  	\label{fig:compl4}
  \end{figure}
As an example, in Figure~\ref{fig:compl4} we show the values of $\delta\omega_{R}$ and $\delta \omega_{I}$ for the modes $(\ell,m,n)=(4,0,0)$, $(4,2,0)$ obtained through the modified Teukolsky equation and those obtained through our eikonal approximation (black lines). The corrections from the modified Teukolsky equation come in two types, denoted $\delta\omega^{\pm}$, and reflect the breaking of isospectrality --- we recall that the theory is only isospectral in the eikonal regime. 
However, we observe that $\delta\omega^{+}$ and $\delta\omega^{-}$ are very close to each other, implying that the theory has almost converged to the isospectral regime already at $\ell=4$. 
In addition, the eikonal prediction is very close to each pair of curves and shows that the eikonal approximation is remarkably good even for relatively low values of $\ell$ and $m$. 

  \begin{figure}[t!]
  	\centering
	\includegraphics[width=0.99\textwidth]{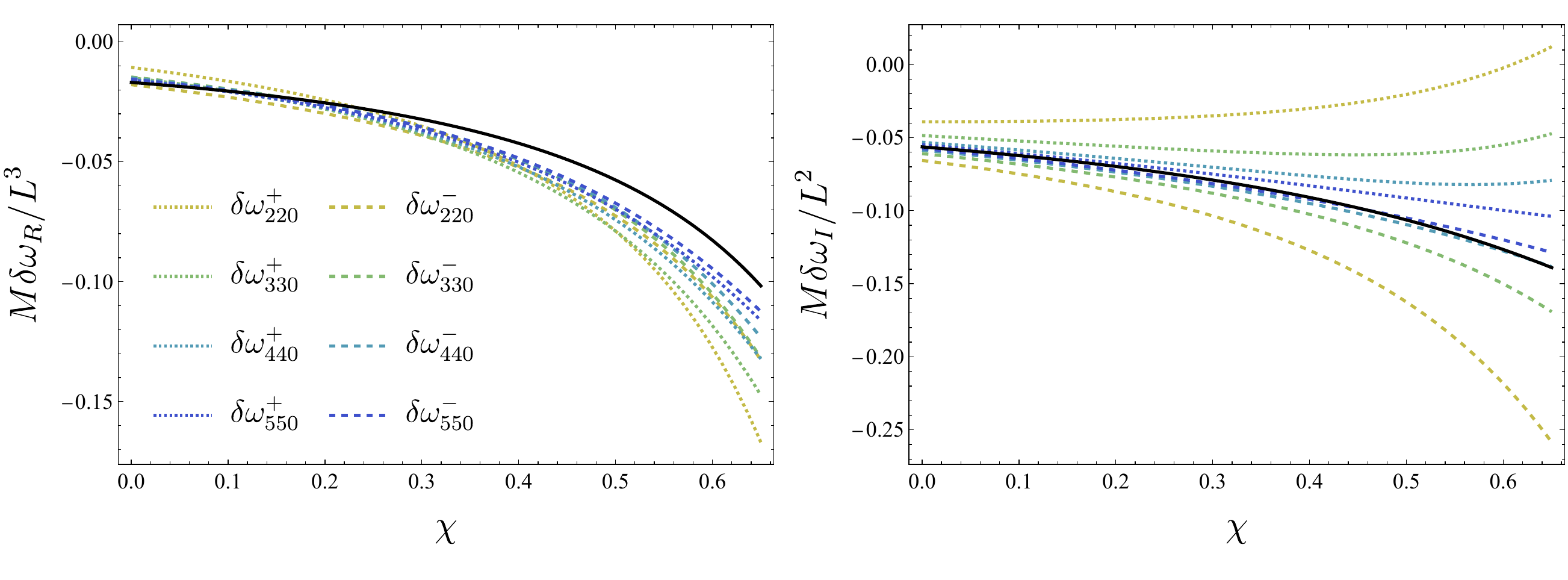}
  	\caption{Convergence towards the eikonal regime for $\ell=m$ modes. 	
	The colored dotted and dashed lines are the prediction from the modified Teukolsky equation for $(\ell,m,n)=(2,2,0), (3,3,0), (4,4,0), (5,5,0)$. The solid black lines are the prediction from our eikonal computation.}
  	\label{fig:convergencelm}
  \end{figure}

We offer a different visualization in Figure~\ref{fig:convergencelm} where we analyze the convergence of the $\ell=m$ fundamental modes towards the eikonal prediction. We have included frequencies up to $\ell=5$, which we have computed following \cite{Cano:2024ezp} and using the resources in \cite{gitbeyondkerr}. In the case of the real part, we can see the convergence is fast and even for $\ell=m=2$ the eikonal result gives a reasonably good approximation. In the case of the imaginary part,  the $\ell=m$ modes seem to show a slower convergence towards the eikonal regime than their $m<\ell$ counterparts in Figure~\ref{fig:compl4}. However,  it is clear from the right panel of Figure~\ref{fig:convergencelm} that they still converge towards the eikonal prediction quite fast as we increase $\ell$. 

This remarkable agreement not only validates our approach and results, but it indicates that the eikonal approximation is quite accurate even for low values of $\ell$ and $m$.

\subsection{Results for arbitrary rotation}\label{sec:arbitraryrotation}
The most interesting aspect of our results is that they allow us to investigate for the first time the corrections to the QNMs for arbitrarily large rotation, including extremality. Here we investigate in detail some of the most interesting features of the QNM spectrum. We start by reviewing some aspects of the Kerr QNMs that are important to understand the behavior of the corrections. 

In the case of Kerr black holes, an important feature is the existence of a ``phase boundary'' at a critical value of $\mu=\mu_{\rm cr}$, which separates modes with different behavior \cite{Yang:2012he,Hod:2012bw,Yang:2012pj,Yang:2013uba}. For $\mu>\mu_{\rm cr}$, the imaginary part of the modes tends to zero in the extremal limit $\chi\to 1$, and thus they are denoted zero-damping modes (ZDMs). In fact, all these modes tend to the special value
\begin{equation}\label{ZDM}
\omega^{\rm Kerr}_{R}=m\Omega_{H}-\frac{L}{4M}\epsilon^{1/2}\sqrt{-8+15\mu^2-\mu^4}+\mathcal{O}(\epsilon)\, ,\quad \omega^{\rm Kerr}_{I}=-\frac{1}{M}\left(n+\frac{1}{2}\right) \sqrt{\frac{\epsilon}{2}}+\mathcal{O}(\epsilon)\, ,
\end{equation}
for $\epsilon=1-\chi\ll 1$, where $\Omega_{H}=1/(2M)$ is the horizon's angular velocity at extremality. 
For $\mu<\mu_{\rm cr}$ the modes tend to a different value with $\omega_{I}^{\rm Kerr}\neq 0$, and thus they are damped modes (DMs). The separation between both behaviors can be understood by looking at the shape of the potential $U$ at extremality. The horizon $r_{+}$ is always an extremum of the potential at extremality for $\omega=m\Omega_{H}$. However, it can be either a maximum or a minimum --- we remark that the WKB method identifies the QNMs in terms of the \emph{minimum} of $U$.\footnote{This means that $\omega^2-U$, which is the usual notion of effective potential, has a maximum.} When $\mu\ge \mu_{\rm cr}$ the horizon is a minimum at extremality; the QNMs identified through the WKB method live on the horizon and they are the ZDMs in \req{ZDM}. On the contrary, when $\mu<\mu_{\rm cr}$ the horizon is a maximum of the potential at extremality. The potential develops a minimum outside the horizon, which is the one captured by the WKB method, and this leads to DMs. The transition between both types of modes happens when the horizon changes from a minimum to a maximum, so that it becomes a saddle point:
\begin{figure}[h!]
  	\centering
  	\includegraphics[width=0.99\textwidth]{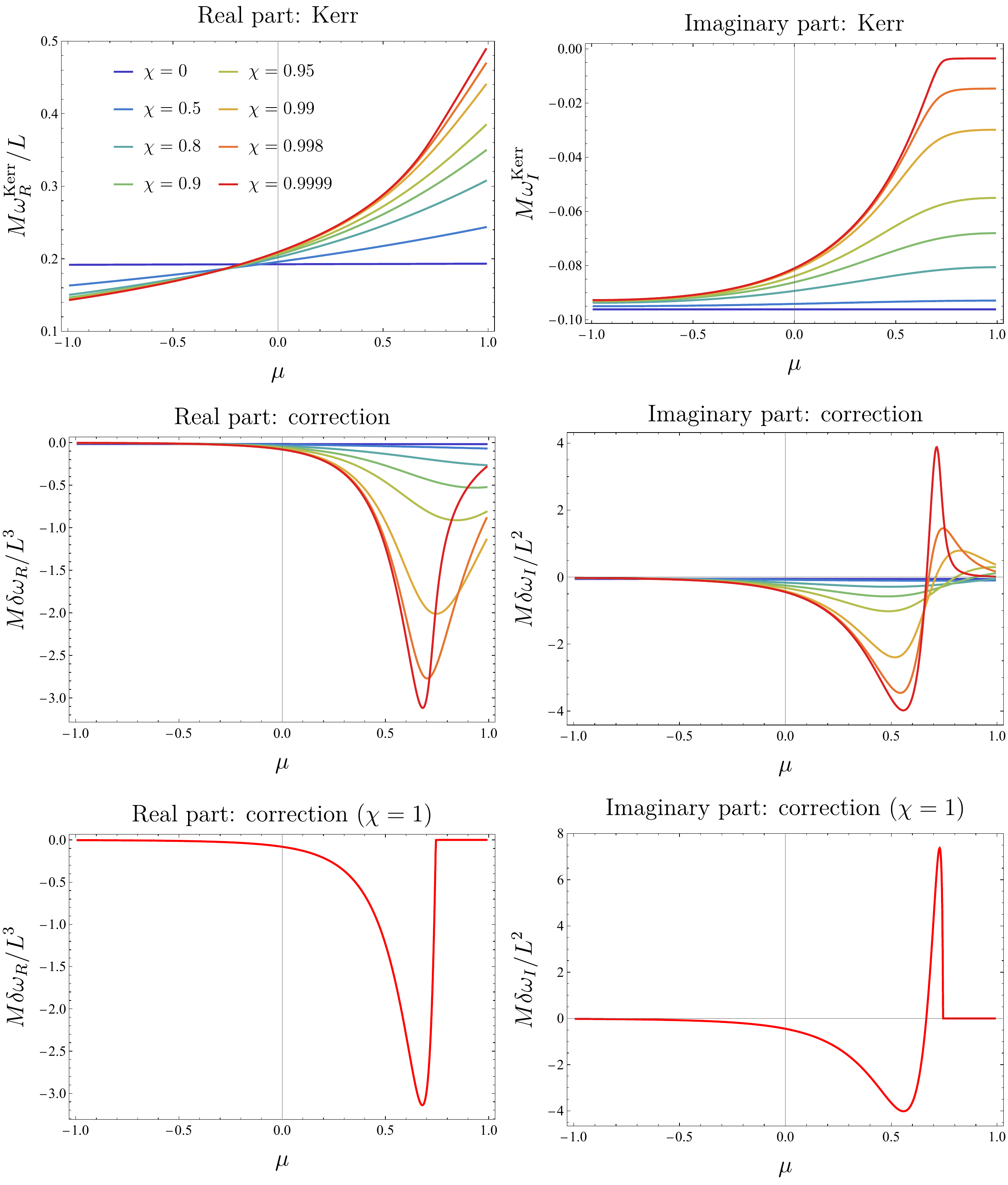}
  	\caption{Kerr QNM frequencies (top row) and their corrections (middle and bottom row) as a function of $\mu=m/L$ for different values of the black hole rotation $\chi$. In the bottom row we show the corrections at extremality. We normalize the frequencies by their scaling with $L$, and the imaginary part corresponds to the fundamental mode $n=0$. }
  	\label{fig:omegamu}
  \end{figure}
\begin{equation}\label{saddle}
\frac{\partial^2 U}{\partial r^2}\bigg|_{\chi=1,\, r=r_{+},\, \omega=m\Omega_{H}}=0\, .
\end{equation}
As shown by \cite{Yang:2012he,Hod:2012bw,Yang:2012pj,Yang:2013uba}, this condition leads to the identification of the critical value of $\mu$:\footnote{The analytic expression is approximate because it uses the approximation \req{Almeik} for the separation constants. It is nevertheless very close to the exact value $\mu_{\rm cr}=0.74398...$ computed in \cite{Yang:2012pj,Yang:2013uba}.}
\begin{equation}\label{mucr}
\mu_{\rm cr}\approx \sqrt{\frac{15-\sqrt{193}}{2}}\approx 0.744\, .
\end{equation}
This value also plays an important role to understand the corrections to the Kerr QNMs that we discuss next. Let us also point out that, as discovered by \cite{Yang:2012pj,Yang:2013uba}, ZDMs also exist for all $0<\mu<\mu_{\rm cr}$, although they are not captured by the WKB method. Thus, our analysis does not consider those modes, whose identification is left for future work.

In Figure~\ref{fig:omegamu}, we plot the Kerr frequencies and the corrections as a function of $\mu$ for different values of the black hole spin $\chi$, including near-extremal values. In the case of the imaginary part of the Kerr QNMs, the transition between DMs and ZDMs at $\mu_{\rm cr}$ can clearly be seen. 
In the case of the corrections, we observe several remarkable features. First, we see that the corrections become much larger as we increase the rotation, and the curves show an important variation from $\chi=0.99$ to $\chi=1$ --- especially the imaginary part. The corrections become especially prominent around the critical value $\mu_{\rm cr}$. At extremality (bottom row in Figure~\ref{fig:omegamu}), the curves peak at $\mu\lesssim \mu_{\rm cr}$ and they abruptly drop to zero for $\mu>\mu_{\rm cr}$, therefore implying that the ZDMs still approach the value $\omega\to m \Omega_{H}$ at extremality. 
The behavior near $\mu\approx \mu_{\rm cr}$ and $\chi\approx 1$ is highly involved and it depends on the direction of the limit. We refer the reader to Appendix~\ref{app:crit} for more details. On the other hand, these plots show that the corrections are very small for counter-rotating modes $\mu<0$. 

\begin{figure}[h!]
  	\centering
  	\includegraphics[width=0.99\textwidth]{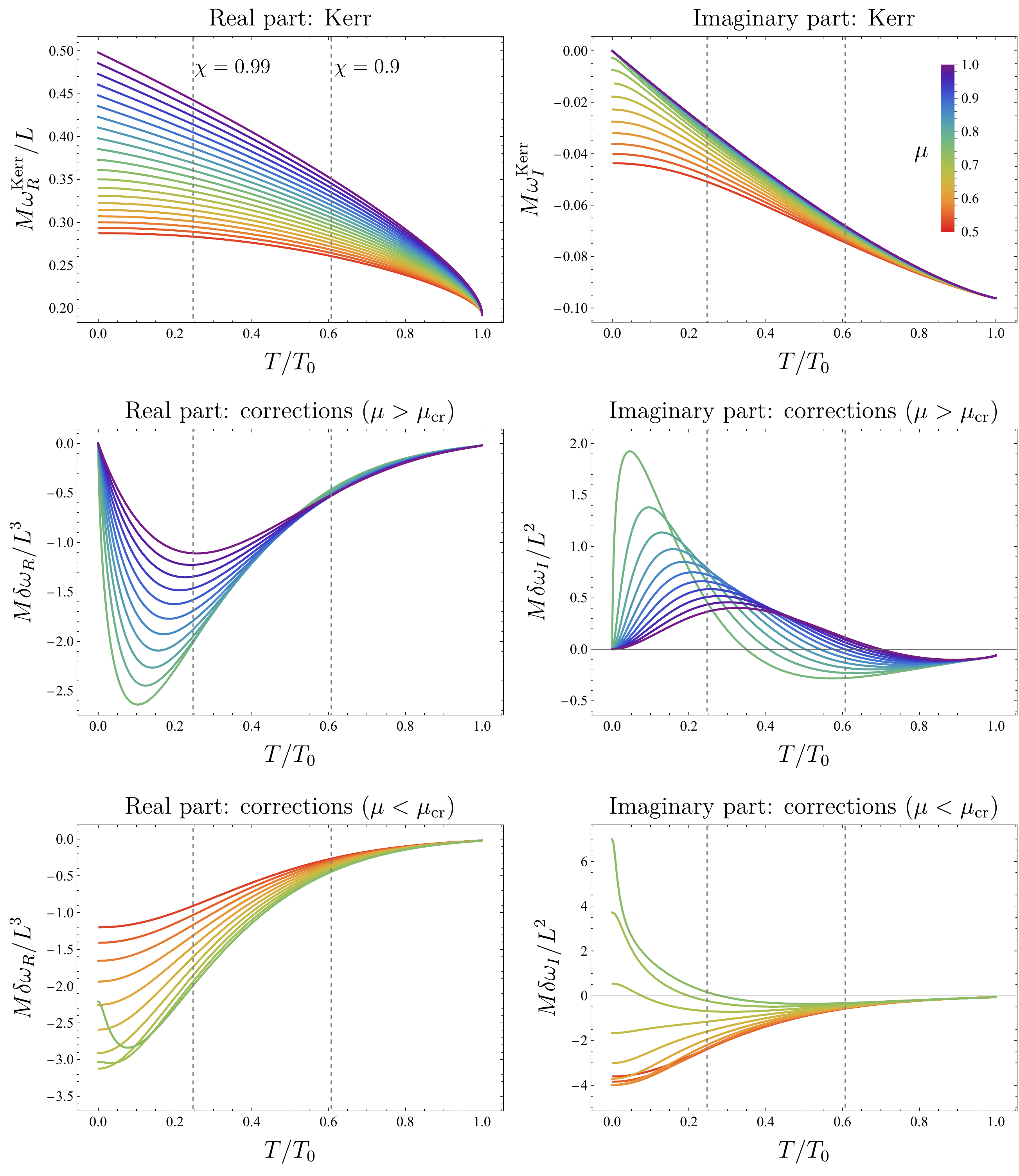}
  	\caption{Kerr QNM frequencies an their corrections as a function of the black hole temperature for different values of $\mu$. In the case of the corrections, we show in different plots the cases of $\mu>\mu_{\rm cr}$ (middle row) and $\mu<\mu_{\rm cr}$ (bottom row). The imaginary part corresponds to $n=0$.}
  	\label{fig:omegaT}
  \end{figure}

In order to obtain more information about the behavior of the QNMs, we can study their dependence on $\chi$ for fixed $\mu$. Since, as we have seen, most of the variation happens for very high rotation, it is useful to work in terms of the temperature
\begin{equation}
T=T_{0} \frac{2\sqrt{1-\chi^2}}{1+\sqrt{1-\chi^2}}\, ,
\end{equation}
where $T_{0}=(8\pi M)^{-1}$ is the temperature at zero rotation. Observe that $T/T_{0}\approx 2\sqrt{2}\sqrt{1-\chi}$ for $\chi\to 1$, so that a quadratically small $1-\chi$ only produces a linearly small $T$, allowing us to zoom in into the near-extremal regime. We show in Figure~\ref{fig:omegaT} the Kerr QNM frequencies and the corrections as a function of $T$ for different values of $\mu>1/2$. For reference, the two vertical lines in each plot mark the cases of $\chi=0.9$ and $\chi=0.99$. As we can see, the greatest part of the variation of $\delta\omega_{R}$ and $\delta\omega_{I}$ occurs for $\chi>0.9$. In absolute value, these shifts can be one order of magnitude larger for high rotation than for moderate rotation $(\chi\sim 0.7-0.8)$.  In relative value, the effects are even more dramatic, and we will come back to this in a moment.

In the case of $\mu>\mu_{\rm cr}$ (middle row in Figure~\ref{fig:omegaT}), the corrections are maximized at some large $(\chi\gtrsim 0.99)$ but sub-extremal value of rotation, and they drop to zero at $T=0$. In fact, an analytical computation shows that $\delta\omega_{R}\propto T$ and $\delta\omega_{I}\propto T^2$ for small $T$. For the real part we find the expression
\begin{equation}
\delta\omega_{R}=-L^3\frac{9(8-7\mu^2+\mu^4)^2}{\sqrt{2}\sqrt{-8+15\mu^2-\mu^4}}C(\mu) 8\pi T+\mathcal{O}(T^2) \, ,
\end{equation}
where 
\begin{equation}\label{Cfunc}
C(\mu)=\frac{1}{K(-q)}\int_{0}^{\pi/2} \frac{du}{\left(1+x_0^2\sin^2 u\right)^{4}\sqrt{1+q \sin^2 u}}\Bigg|_{a=M, \,\omega=L\mu/(2M)}\, ,
\end{equation}
and we recall that $x_0$ and $q$ are given by \req{x0} and \req{qsol}.  The expression for $\delta\omega_{I}$ is more cumbersome and not particularly illuminating.  We observe that the validity of these approximations is restricted to smaller and smaller $T$ as we approach $\mu_{\rm cr}$.  

For $\mu<\mu_{\rm cr}$, the corrections do not tend to zero at extremality and usually are maximized there. The corrections are larger for $\mu$ closer to the critical value, and they become more moderate as we decrease $\mu$. 

\begin{figure}[t!]
  	\centering
  	\includegraphics[width=0.8\textwidth]{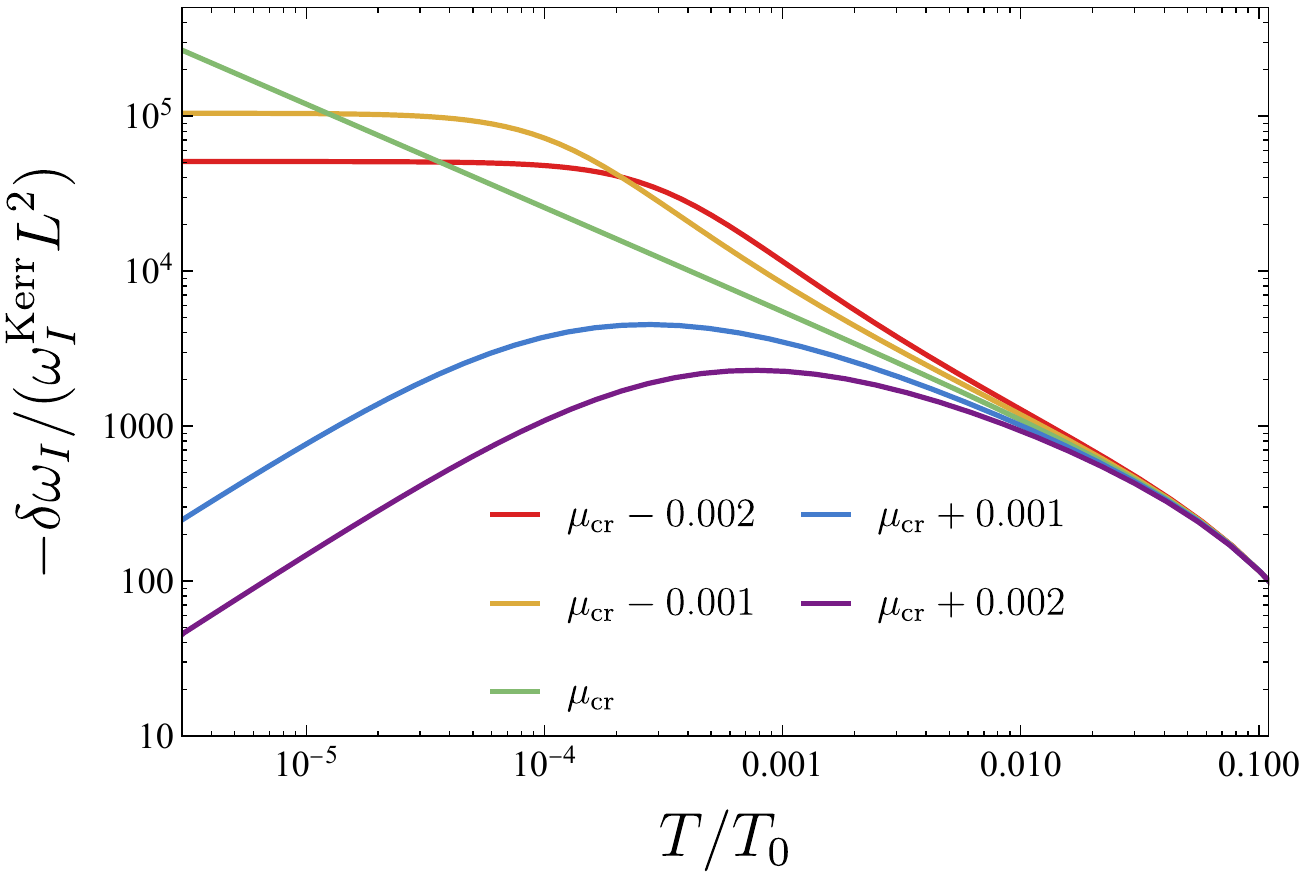}
  	\caption{Relative corrections to the imaginary part of the Kerr QNM frequencies for small temperature and for values of $\mu$ close to the critical value. The corrections become very large in this regime and diverge as $T^{-2/3}$ for $\mu=\mu_{\rm cr}$.}
  	\label{fig:omegaimrel}
  \end{figure}

The relative corrections to the imaginary part --- that is, $\delta\omega_{I}/\omega_{I}^{\rm Kerr}$ --- show an especially intriguing behavior.  It is clear from Figures~\ref{fig:omegamu} and \ref{fig:omegaT}  that, for $\mu$ close to the critical value and $\chi$ close to $1$, the corrections to the imaginary part are maximized at the same time as the imaginary part of the Kerr frequency becomes tiny. This leads us to consider the following question: can the relative correction $\delta\omega_{I}/\omega_{I}^{\rm Kerr}$ grow unboundedly for $\mu\to\mu_{\rm cr}$ and $\chi\to 1$? As it turns out, the answer is affirmative. 
We illustrate this in Figure~\ref{fig:omegaimrel}, where we show $\delta\omega_{I}/\omega_{I}^{\rm Kerr}$ for small $T$ and small $|\mu-\mu_{\rm cr}|$. The green line corresponds to $\mu=\mu_{\rm cr}$ and the numerical results indicate that it grows as $T^{-2/3}$ when $T\to 0$ (thus it shows as a straight line in the log-log plot). One can in general achieve arbitrarily large values of $\delta\omega_{I}/\omega_{I}^{\rm Kerr}$ by taking $\mu$ sufficiently close to $\mu_{\rm cr}$ --- either from above or from below --- and $T$ sufficiently close to $0$.

These findings can be confirmed by an analytic computation.  As we show in Appendix~\ref{app:crit}, the approach to $\mu\to \mu_{\rm cr}$ and $T\to 0$ is different depending on the relative size of $|\mu-\mu_{\rm cr}|$ and $T/T_{0}$. The regime that is most interesting for us is the one corresponding to 
\begin{equation}\label{regime}
|\mu-\mu_{\rm cr}|^3\ll \frac{T^2}{T_{0}^2}\ll 1\, .
\end{equation}
In this case, we find the following approximate expressions for the imaginary part of the Kerr frequency and its correction
\begin{align}
	\omega_{I}^{\rm Kerr}&=-\left(n+\frac{1}{2}\right)\sqrt{6}\pi T\, ,\\
	\delta\omega_{I}&=\left(n+\frac{1}{2}\right)192\sqrt{6} \pi L^2\mu_{\rm cr}^2C(\mu_{\rm cr}) T^{1/3}T_{0}^{2/3}\, ,
\end{align}
with $C(\mu_{\rm cr})\approx 0.52$.  We see that $\omega_{I}^{\rm Kerr}$ goes to zero faster than $\delta\omega_{I}$, implying that, for small enough $T$, the relative size of the correction can be arbitrarily large
\begin{equation}\label{reldelta}
	\frac{\hat{\alpha}\delta\omega_{I}}{\omega_{I}^{\rm Kerr}}=-192 \hat{\alpha}\mu_{\rm cr}^2 C(\mu_{\rm cr})L^2 \left(\frac{T}{T_{0}}\right)^{-2/3}\approx -55.5 \hat{\alpha} L^2 \left(\frac{T}{T_{0}}\right)^{-2/3}\, .
\end{equation}
In principle, the correction can become of order 1 sufficiently close to extremality, namely, when
\begin{equation}
	1-\chi\lesssim  2.1\times 10^{4} L^{6}\hat{\alpha}^3\, .
\end{equation}
However, there is a catch, since \req{reldelta} is only valid in the regime \req{regime}. Thus, if we want  $T/T_{0}$ to take smaller values, we should also decrease $|\mu-\mu_{\rm cr}|$. The interesting observation is that, since $\mu$ is actually a rational number, $\mu=m/L$, we cannot just take it arbitrarily close to $\mu_{\rm cr}$, which is irrational.\footnote{Here we are using the approximation \req{mucr} for $\mu_{\rm cr}$, which is irrational. Whether the exact value, identified in \cite{Yang:2012pj,Yang:2013uba}, is also an irrational number, is probably a difficult mathematical question. However, the most likely possibility is that it is indeed irrational.} Intuitively, for a fixed $L$ (we recall it is a semi-integer), the best approximation to $\mu_{\rm cr}$ that we can obtain with a rational number is $|\mu-\mu_{\rm cr}|\sim 1/L$. However, Dirichlet's approximation theorem tells us that we can do better. This theorem implies the following. If $L$ is a large number  and we consider fractions $m/L'$, with $L'$ of the same order of magnitude as $L$, then the approximation can in general be improved to $|\mu-\mu_{\rm cr}|\sim 1/L^2$. Therefore, how large can the relative corrections realistically get? If we take into account \req{regime} and use $|\mu-\mu_{\rm cr}|\sim 1/L^2$, we conclude that the minimum temperature that we can consider for a given $L$ is $T/T_0\gtrsim L^{-3}$. Plugging this into \req{reldelta}, we find the following bound on the maximum size of the corrections
\begin{equation}\label{reldelta2}
\left|\frac{\hat{\alpha}\delta\omega_{I}}{\omega_{I}^{\rm Kerr}}\right|\lesssim |\hat{\alpha}| L^4 \, .
\end{equation}
Finally, we recall that $L$ cannot be arbitrarily large, as this would take us away from the regime of validity of the EFT.  Assuming the upper bound in \req{eq:validity}, the maximum value that $L$ can take is $\to L\lesssim |\hat\alpha|^{-1/6}$, which implies that the maximum size of the relative corrections is
\begin{equation}\label{reldelta3}
\left|\frac{\hat{\alpha}\delta\omega_{I}}{\omega_{I}^{\rm Kerr}}\right|\lesssim |\hat\alpha|^{1/3} \, .
\end{equation}
Although they cannot overcome the GR prediction while respecting the EFT requirements, this is still a remarkable result. While usually, the corrections to GR are of order $|\hat\alpha|$, we have found an special regime in which they can grow up to order $|\hat\alpha|^{1/3}$. Since $|\hat\alpha|$ is assumed to be a very small number, this means that there is an increase of many orders of magnitude in the size of the corrections. Furthermore, this bound depends on assuming the most conservative bounds on the regime of validity of the EFT and on assuming $|\mu-\mu_{\rm cr}|\sim 1/L^2$. Interestingly, if it turned out that $\mu_{\rm cr}$ is extremely close to a fraction, then the corrections could be even larger. \\

Another important question is that about the interpretation of the divergence in \req{reldelta} (assuming that $\mu$ could be taken to be exactly $\mu_{\rm cr}$). This is addressed in detail in Ref.~\cite{Cano:2025ejw}, where we show that this divergence is not a breakdown of the EFT expansion, but rather a breakdown of the linear-in-$\alpha$ expansion near the critical point $\mu_{\rm cr}$. Therefore, the divergence does signal a legitimate growth of the effects of the higher-derivative corrections \cite{Cano:2025ejw}.

We can give an intuition about why the modes around $\mu_{\rm cr}$ can be expected to be very sensitive to corrections. The reason is that higher-derivative corrections modify the boundary between ZDMs and DMs, and thus the \emph{actual} value of $\mu_{\rm cr}$ is modified.  In fact, when we take into account the higher-derivative corrections to the potential, the condition \req{saddle} becomes 
\begin{equation}
\frac{16M^4}{L^2}\frac{\partial^2 U}{\partial r^2}\bigg|_{\chi=1,\, r=r_{+},\, \omega=m\Omega_{H}}=-8+15\mu^2-\mu^4+144L^2\hat{\alpha} (8-7\mu^2+\mu^4)^2 C(\mu)=0\, ,
\end{equation}
where $C(\mu)$ is given by \req{Cfunc}. Solving this equation as an expansion in $\hat\alpha$ yields
\begin{equation}\label{mucr2}
\mu_{\rm cr}\approx \mu_{\rm cr}^{(0)}-71.3 \hat{\alpha} L^2\, , 
\end{equation}
where $\mu_{\rm cr}^{(0)}$ is the original value in \req{mucr}. This implies that the modes located in the band between $\mu_{\rm cr}$ and $\mu_{\rm cr}^{(0)}$, change their behavior. If $\alpha>0$, the modes in this band were originally DMs for the Kerr case, and they become ZDMs in the higher-derivative theory. If $\alpha<0$, the opposite situation happens: modes that were originally ZDMs transition into DMs. This is a large modification of the spectrum, since the lifetimes of the modes are dramatically modified in the extremal limit. Therefore, the large corrections to the imaginary part in \req{reldelta} can be attributed to this change in the nature of the QNMs. 

Now, in reality $\mu$ is not actually continuous; it is a fraction.  Hence, we face the same question as before: can we actually find values of $\mu=m/L$ between  $\mu_{\rm cr}$ and $\mu_{\rm cr}^{(0)}$ while remaining in the regime of validity of the EFT? This is equivalent to asking whether there are modes such that
\begin{equation}
\left|\mu-\mu_{\rm cr}^{(0)}\right|< 71. 3 |\hat \alpha| L^2\, .
\end{equation}
As we have seen earlier, in general we can find many modes for which $\left|\mu-\mu_{\rm cr}^{(0)}\right|\sim 1/L^2$. However, this is not small enough, since \req{eq:validity} implies $1/L^2\gg |\hat{\alpha}| L^4\gg  |\hat{\alpha}| L^2$. This means that, for the great majority of the modes, the corrections are not large enough to make them cross the boundary between ZDMs and DMs. 
Despite this, it cannot be discarded that, by accident, there is an extremely good rational approximation to $\mu_{\rm cr}^{(0)}$ with a low denominator. This would imply the existence of modes that lie extremely close to the phase boundary and could therefore cross the boundary when the corrections are included. To investigate this question properly, we would need to analyze the modified Teukolsky equation for finite values of $L$ (not only the eikonal limit), and the corresponding condition for the phase boundary \cite{Cano:2025ejw}. 
Regardless of whether such modes exist, the modification of the boundary between ZDMs and DMs explains why modes around $\mu_{\rm cr}$ will in general suffer larger corrections near extremality.

\section{Conclusions}\label{sec:conclusions}
To the best of our knowledge, our results represent the first computation of gravitational QNMs of black holes with high rotation --- including extremality --- in an extension of GR.  Although our computation is restricted to the eikonal limit, the good agreement with the results from the modified Teukolsky equation for moderate rotation (see section~\ref{subsec:comparison}) indicates that our results likely provide a good approximation even for low values of $\ell$ and $m$. 

Our analysis has taken advantage of the remarkable properties of the theory \req{eq:isoeft}, which preserves isospectrality in the eikonal limit. We have shown that large momentum perturbations in this theory can be described by an effective scalar equation with higher-derivative corrections \req{effectivePhieq}, in the same way that the wave equation for a scalar field describes eikonal gravitational perturbations in GR. The existence of this universal scalar equation governing gravitational perturbations of arbitrary parity is a manifestation of the isospectrality of the theory. 

We have then  obtained the black hole QNMs from two different approaches. 
On the one hand, we have analyzed the geometric optics limit of the master equation, which allows us to identify QNMs in terms of the properties of the graviton-sphere --- the surface of closed unstable GW orbits around the BH.  These orbits are neither null nor geodesic due to the modification of the dispersion relation, and thus the graviton-sphere is different from the usual photon-sphere. By focusing on the case of equatorial orbits, which correspond to QNMs with $\ell=m$, we identified the real part of the QNM frequency with the orbital frequency of those orbits. In addition, we found that the imaginary part is no longer proportional to the Lyapunov exponent of the circular unstable GW orbits.  This seems to be a consequence of the master equation containing higher-order derivatives of the field.  For non-equatorial orbits the analysis from the geometric optics approach becomes more challenging since the corresponding Hamilton-Jacobi equation is non-separable. 

To analyze the general case $\ell\neq m$ we resorted to a direct analysis of the master scalar equation and we dealt with non-separability by projecting it onto the spheroidal harmonics and finding a decoupled master radial equation.  One of the main accomplishments of our paper is the explicit expression for the correction to the effective potential in \req{deltaV1}. By solving the master radial equation through the WKB approach, we then managed to obtain the perturbative corrections to the Kerr QNMs with arbitrary $\mu=m/(\ell+1/2)$ and arbitrary rotation $\chi$. 

Our results offer a clear headline: the corrections to the QNM spectrum become much larger for high rotation. They furthermore exhibit a rich and complex dependence on $\mu$ and $\chi$. For counter-rotating modes $\mu<0$, the corrections remain small for all values of rotation. For $\mu>0$ they are highly non-linear and grow very rapidly for $\chi$ close to extremality. The effects are especially dramatic when $\mu$ is close to the critical value $\mu_{\rm cr}\approx 0.744$, which represents the boundary between DMs and ZDMs. In fact, we discussed in detail the behavior of QNMs near $\mu_{\rm cr}$ and we discovered that the relative corrections to the imaginary part of the QNM frequencies can in principle become arbitrarily large. The maximum size of the corrections depends on how close we can take $\mu$ --- which is a rational number --- to the critical value $\mu_{\rm cr}$. 
Even under conservative assumptions, we showed that these findings imply the existence of many modes which receive corrections orders of magnitude larger than expected.  On the other hand, if by chance there was an extremely good approximation to $\mu_{\rm cr}$ with a rational number involving a sufficiently low denominator, this could lead to even order-one corrections to GR and large modifications of the spectrum. It is quite amusing that the possibility of having large corrections to GR depends on a question about number theory --- how well a (presumably) irrational number $\mu_{\rm cr}$ can be approximated by a fraction. 
We expect that the sensitivity of QNMs near $\mu_{\rm cr}$ to new physics is a general phenomenon, since the corrections to GR can change the boundary between DMs and ZDMs, and thus can potentially change the nature of the modes lying very close to that boundary. We have argued that the crossing of the boundary cannot take place generically within the regime of validity of the EFT. But again, if there was an extremely good rational approximation to $\mu_{\rm cr}$, there would be modes extremely close to the boundary whose nature could be affected by the corrections. 

Regardless of the existence of special modes that could undergo order-one corrections, our results show that the growth of the corrections near extremality is generic, and this has implications too for astrophysics. If we use our eikonal prediction to estimate QNMs relevant for black hole spectroscopy (Table~\ref{table3}), we see that the corrections can easily be two orders of magnitude larger for high rotation $\chi\sim 0.99-0.999$ than for moderate rotation $\chi\sim 0.7$ --- which is the typical value for most post-merger black holes. This, together with the fact that highly-rotating black holes yield a long-lived ringdown signal that would allow for much more precise measurements, implies that highly-rotating black holes are far superior in order to test physics beyond GR. In fact, in the light of our results, these black holes could be the only events in which it makes sense to look for new physics.  

%\bgroup
%\def\arraystretch{1.24}
%\setlength{\tabcolsep}{4pt}
\begin{table}[h!]
	\centering
	\begin{tabular}{|c||c|c|c|}
		\hline
		$\chi$/$(\ell,m)$ & (2,2)& (3,2)& (4,3)\\
		\hline
		$0.7$& (-3.07, 10.9)& (-4.68, 24.7)& (-8.70, 40.9)\\
		$0.99$& (-32.7, -155)& (-53.4, 616)& (-113, 434)\\
		$0.998$& (-35.6, -450)& (-70.6, 1166)& (-161, 448)\\ \hline
	\end{tabular}
	\caption{Relative corrections $(\delta\omega_{R}/\omega_{R}, \delta\omega_{I}/\omega_{I})$ for selected values of $(\ell, m)$ and the black hole spin $\chi$, estimated from the eikonal prediction with $\mu=m/(\ell+1/2)$. The values may be inaccurate since we are using the eikonal formula outside its regime of validity, but they illustrate the growth of corrections near extremality.}
	\label{table3}
\end{table}
%\egroup

Our work opens several natural directions. As we noted, our approach based on the WKB method does not capture zero-damping modes with $\mu<\mu_{\rm cr}$, which could in turn be obtained via matched asymptotic expansions. It would be important to analyze these modes in order to obtain the complete QNM spectrum and understand the branching of the QNMs \cite{Yang:2012pj,Yang:2013uba}. in the presence of higher derivative corrections. We also focused on perturbative corrections to the QNM frequencies, but it would be interesting to understand if large modifications of the spectrum could take place due to the potential generating a double peak or a well, as depicted in Figure~\ref{fig:potential}. In particular, one would need to determine if those situations can take place within the regime of validity of the EFT. 
On the other hand, it should be possible to extend the correspondence between QNMs and GW orbits beyond the equatorial case. To this end, one should devise a way to deal with the non-separability of the Hamilton-Jacobi equation. 
Finally, it would be interesting to extend the analysis of eikonal QNMs to general EFT corrections. This is more involved than the case of the isospectral theory, since in general we lack a single effective master equation to describe the perturbations. In turn, one would have to decompose the perturbations according to their parity type and find the corresponding dispersion relation or master equation in each case. 

The ultimate goal is to understand the corrections to the QNM spectrum of highly-rotating black holes for the low-$\ell$ modes as well. This is still challenging, as no existing method seems adequate to address it, but our findings provide a further motivation to pursue this goal.

\section*{Acknowledgements}
We would like to thank Alfredo Gonz\'alez Lezcano and Matthew Roberts for useful discussions.
The work of PAC received the support of a fellowship from “la Caixa” Foundation (ID 100010434) with code LCF/BQ/PI23/11970032. MD is supported by the Postdoctoral Fellows of the Research Foundation - Flanders grant (1235324N). GvdV is supported by a Proyecto de Consolidación Investigadora (CNS2023-143822) from Spain’s Ministry of Science, Innovation and Universities. PAC and MD would also like to thank the hospitality of Benasque Science Center where the final part of this work was completed. This research was carried out in part during a scientific visit funded by the COST action CA22113 "Fundamental Challenges in Theoretical Physics."

\appendix

\section{Expansion near the critical point}\label{app:crit}

In this Appendix, we study the QNM frequencies near the critical point for black holes with large rotation. These modes satisfy $\vert \mu-\mu_{\text{cr}}\vert \ll 1$ and are near-extremal $\epsilon \equiv 1-\chi \ll 1$. Although both quantities are small, it is crucial to keep note of their relative sizes when expanding, which we now hope to clarify. We can then straightforwardly find both the real and imaginary part of the Kerr frequencies, by substituting the analytic expansion of $r_0^{\text{Kerr}}$ in \eqref{wRKerr} and \eqref{wIKerr}, respectively. The approximate expressions for the corrections can be obtained by the same token, using \eqref{r0gen}, \eqref{deltawRgen} and \eqref{deltawIgen}.

More concretely, from (\ref{zchirel}) we can solve for $\chi$ as a function of $z$. This is a quite cumbersome expression, but around $z=1$, we can solve it perturbatively
\begin{equation} \label{eq:eqforchi}
	\chi = 1 + \frac{8 - 15\mu^2 + \mu^4}{16 \mu^2}(z-1)^2 - \frac{16 - 22\mu^2 + 3\mu^4 + \mu^6}{16 \mu^4}(z-1)^3 + \dots \,.
\end{equation}
We can expand once more around $\mu=\mu_c$ to find the simpler expression
\begin{equation}\label{chi(z)}
	\chi\approx 1-\eta (\mu-\mu_c)(z-1)^2-(z-1)^3 + \dots \, ,
\end{equation}
where $\eta \equiv \frac{\sqrt{193}}{8 \mu_c}\approx 2.33$ and the dots indicate higher order terms in the expansion. Technically, it is enough to just retain the quadratic term in $(z-1)$ and truncate the higher order contributions in \eqref{chi(z)}. However, as we approach the critical value of $\mu$, the quantity $|\mu-\mu_c|$ may be become smaller than $\epsilon$. Then the quadratic term can become subleading relative to the cubic term in $(z-1)$. Because of this competition, both terms should be kept and we must analyze the roots of \eqref{chi(z)} carefully. As a first step, we can implement a linear transformation so that the equation no longer has quadratic terms in $z$. In this form, we arrive at a depressed cubic equation and the roots $z_k$ can be written more compactly using trigonometric functions
\begin{equation}
	z_k = 1 - \frac{A}{3B} + \frac{2A}{3B} \cos\left(\frac{1}{3} \arccos\left[-1 + \frac{27 B^2 \epsilon}{2 A^3}\right] + \frac{2\pi k}{3} \right)\,,
\end{equation}
where
$k=0,1,2$ is the labeling of the roots and
\begin{align}
	A = -\frac{8 - 15\mu^2 + \mu^4}{16 \mu^2} \approx \eta (\mu-\mu_c), \quad B= \frac{16 - 22\mu^2 + 3\mu^4 + \mu^6}{16 \mu^4} \approx 1\,.
\end{align}
Note that $\epsilon$ only appears in the argument of the arc-cosine. It is now clear that the expansion parameter that is sensitive to the relative sizes of $\epsilon$ and $\delta \mu \equiv \mu-\mu_c$ is given by
\begin{align}
	\sigma = \frac{\epsilon}{A^3} = \frac{\epsilon}{\eta^3 \delta \mu^3}.
\end{align}
The function \eqref{eq:eqforchi} is qualitatively different depending on the sign of $\delta\mu$, which reflects the presence of a phase boundary at $\mu=\mu_{\text{cr}}$. If $\delta\mu>0$, the horizon $z=1$ is a maximum and we must take the $k=0$ solution. On the other hand, if $\delta\mu<0$ it is a local minimum and the solution corresponding to $k=1$ must be considered. The critical case $\delta\mu=0$ corresponds to a saddle point at $z=1$.
\begin{table}
\begin{center} 
	\renewcommand{\arraystretch}{2.0} % Local vertical spacing
	\begin{tabular}{|m{6cm}|>{\centering\arraybackslash}m{2cm}|>{\centering\arraybackslash}m{5cm}|}
		\hline
		\centering\textbf{Expansion} & \textbf{Regime} & \textbf{Value of $\delta \mu$} \\
		\hline
		$\displaystyle z - 1 \approx \frac{\epsilon^{1/2}}{A^{1/2}} - \frac{B}{2A^2}\epsilon + \ldots$
		& $\ |\sigma| \ll 1$ & for any $\delta \mu > 0 $ \\
		\hline
		$\displaystyle z - 1 \approx \epsilon^{1/3} - \dfrac{1}{3} \eta \delta \mu + \ldots$
		& $|\sigma| \gg 1$ & for any $\mu$ satisfying $|\delta \mu| \ll 1$ 
		\\ \hline
		$\displaystyle z - 1 \approx -\eta \delta \mu + \dfrac{\epsilon}{\eta^2 \delta \mu^2} + \ldots$
		& $|\sigma| \ll 1$ & $\delta \mu < 0,\ |\delta \mu| \ll 1$ \\ 
		\hline
	\end{tabular}
\end{center}
\caption{The different expansions of $(z-1)$ for the two possible regimes.}
\label{table:expansions}
\end{table}
In Table~\ref{table:expansions}, we summarize the different value of $z-1$ depending on the size of $\sigma$ and the sign and size of $\delta \mu$. It is clear that the expansion of the QNM frequencies will also depend on which regime we choose. Stated differently, the order of limits between extremality and criticality plays an important role in the expansion. In the following, we present the QNMs for the three distinct regimes.
For $|\sigma| \ll 1$ and any value of $\delta \mu >0$, i.e., the first line in Table~\ref{table:expansions}, we find
\begin{subequations}
\begin{align} 
	\frac{M \omega_{\text{R}}^{\text{Kerr}}}{L}&= \frac{\mu }{2}-\mu \sqrt{A \epsilon}
	+\ldots \,,
	\\
	\frac{M \omega_{\text{I}}^{\text{Kerr}}}{(n+1/2)}&= - \sqrt{\frac{\epsilon}{2}}
	+ \ldots \,,
	\\
	\frac{M\delta\omega_{\rm R}}{L^3}&= -\frac{9}{2\mu}\left(\mu ^4-7 \mu ^2+8\right)^2 C\left(\mu_{}\right)\sqrt{ \frac{\epsilon}{A}} 
	+ \ldots \,,
	\\
	\frac{M \delta\omega_{\text{I}}}{L^2(n+1/2)}&= \mathcal{O}(\epsilon)\,.
\end{align}
\end{subequations}
The second regime, $|\sigma| \gg 1$ and $\vert \delta \mu\vert \ll1$ (the second row in Table~\ref{table:expansions}), is more interesting as it allows us to continuously connect to exact criticality $\mu=\mu_{\text{cr}}$. Assuming this hierarchy of scales, we have
\begin{subequations}
\begin{align} 
	\frac{M \omega_{\text{R}}^{\text{Kerr}}}{L}&= \frac{\mu}{2} - \frac{3}{4}\mu \epsilon^{2/3} +\ldots \,,
	& \quad
	\frac{M \omega_{\text{I}}^{\text{Kerr}}}{(n+1/2)}&= -\frac{\sqrt{3}}{2}\epsilon^{1/2}+\ldots \,,
	\\
	\frac{M\delta\omega_{\rm R}}{L^3}&=- 288 
	\mu_{\rm cr}^3C\left(\mu_{\rm cr}\right)\epsilon^{1/3}+\ldots \,,
	& \quad
	\frac{M \delta\omega_{\text{I}}}{L^2(n+1/2)}&= 48 \sqrt{3} \mu_{\rm cr}^2 C\left(\mu_{\rm cr}\right) \epsilon^{1/6}+\ldots \,,
\end{align}
\end{subequations}
where $C(\mu)$ is defined in (\ref{Cfunc}). Taking into account that the temperature goes like $T/T_0\approx 2\sqrt{2}\epsilon^{1/2}$ near extremality, we can easily derive equation \eqref{reldelta}, which implies an enhancement of the relative correction for the imaginary part at the critical point.

Finally, for $|\sigma| \ll 1$ and $\delta \mu \ll 1$ but $\delta\mu < 0$, we find
\begin{subequations}
\begin{align} 
	\frac{M \omega_{\text{R}}^{\text{Kerr}}}{L}&= \frac{\mu}{2} + \frac{\delta \mu ^2 \eta ^2 \mu}{4} +\ldots \,,
	\quad &
	\frac{M \omega_{\text{I}}^{\text{Kerr}}}{(n+1/2)}&= -\frac{(-\delta \mu \eta)^{3/2}}{2}+\ldots \,,
	\\
	\frac{M\delta\omega_{\rm R}}{L^3}&= 288 \mu_c^3  C\left(\mu_{\rm cr}\right) \eta \delta \mu +\dots  \,,
	\quad &
	\frac{M \delta\omega_{\text{I}}}{L^2(n+1/2)}&= 432 \mu_{\rm cr}^{2} C\left(\mu_{\rm cr}\right) \sqrt{- \eta \delta \mu }   + \ldots \,.
\end{align}
\end{subequations}

	%\renewcommand{\leftmark}{\MakeUppercase{Bibliography}}
	%\phantom-section
	\bibliographystyle{JHEP}
	\bibliography{Gravities.bib}
	%\label{biblio}
	
\end{document}